%% file: aa.tex
\documentclass[12pt]{report} 
\usepackage{epsf}
\usepackage{natbib}
\usepackage[sectionbib]{bibunits}
\usepackage{graphicx}
\usepackage{lscape}
\usepackage{rotating}
\usepackage{times}
\usepackage{titlesec}
\input{define.tex}

\begin{document}
	\eject

\linespread{1.5}
\selectfont 
%\special{!userdict begin /bop-hook{gsave 200 30 translate 66 rotate
 % /Times-Roman findfont 100 scalefont setfont 0 0 moveto 0.93 setgray
  %(\today) show grestore}def end}
\pagestyle{empty}
	\include{title}
	\setcounter{page}{1}
	\linespread{1.5}
	\selectfont 
	\pagestyle{plain}
	\renewcommand{\thepage}{\Roman{page}}
%	\tableofcontents
%	\listoffigures
%	\listoftables
%	\eject
	\newpage
	\renewcommand{\thepage}{\arabic{page}}
	\setcounter{chapter}{0}
%Headers and footers
	\pagestyle{plain}
%	\rhead{\nouppercase{\rightmark}}	
%	\lhead{\nouppercase{\leftmark}}
	\renewcommand{\headsep}{15pt}
	\setcounter{page}{1}
\bibliographyunit[\section]
\bibliographystyle{mn2e}
	       	\include{chapter_Introduction}
		\include{chapter_Science}

		\pagestyle{empty}
		\pagestyle{plain}
	\pagestyle{empty}
\end{document}

%% file: define.tex
\newcommand{\aap}{{\it A\&A~\/}}
\newcommand{\aaps}{{\it A\&AS~\/}}
\newcommand{\aj}{{\it AJ~\/}}
\newcommand{\apj}{{\it ApJ~\/}}
\newcommand{\apjl}{{\it ApJL~\/}} 
\newcommand{\apjs}{{\it ApJS~\/}}
\newcommand{\araa}{{\it ARA\&A~\/}}
\newcommand{\mnras}{{\it MNRAS~\/}} 
\newcommand{\apss}{{\it Astrophys. Space. Sci.~\/}}
\newcommand{\pasp}{{\it PASP~\/}}
\newcommand{\nat}{{\it Nature~\/}}
\def\H0{{\rm ~km~s^{-1}~Mpc^{-1}}}

\def\la{\mathrel{\hbox{\rlap{\hbox{\lower4pt\hbox{$\sim$}}}{\raise2pt\hbox{$<$}}
}}}
\def\ga{\mathrel{\hbox{\rlap{\hbox{\lower4pt\hbox{$\sim$}}}{\raise2pt\hbox{$>$}}
}}}
\def\d25{D$_{25}$}

\newbox\grsign \setbox\grsign=\hbox{$>$} \newdimen\grdimen
\grdimen=\ht\grsign
\newbox\simlessbox \newbox\simgreatbox \newbox\simpropbox
\setbox\simgreatbox=\hbox{\raise.5ex\hbox{$>$}\llap
     {\lower.5ex\hbox{$\sim$}}}\ht1=\grdimen\dp1=0pt
\setbox\simlessbox=\hbox{\raise.5ex\hbox{$<$}\llap
     {\lower.5ex\hbox{$\sim$}}}\ht2=\grdimen\dp2=0pt
\setbox\simpropbox=\hbox{\raise.5ex\hbox{$\propto$}\llap
     {\lower.5ex\hbox{$\sim$}}}\ht2=\grdimen\dp2=0pt

\def\simless{\mathrel{\copy\simlessbox}}

%% file: title.tex
\begin{center}
$\: $ \\
\vspace{2cm}

{\huge  {\bf White Paper}}\\
\vspace{1cm}
{\huge {\bf GAIA and the end states of stellar evolution}}\\
\vspace{3cm}

{\large Editors: M.A.Barstow$^{1}$, S.L. Casewell$^{*1}$, B. Gaensicke$^{2}$, J. Isern$^{3}$, S. Jordan$^{4}$}\\

\vspace{1cm}

{\normalsize Authors: M.A~Barstow$^{1}$, S.L.~Casewell$^{1}$, S.~Catalan$^{2}$, C.~Copperwheat$^{5,2}$,
  B.~Gaensicke$^{2}$, E.~Garcia-Berro$^{6}$, N.~Hambly$^{7}$, U.~Heber$^{8}$, J.~Holberg$^{9}$, J.~Isern$^{3}$,
  S.~Jeffery$^{10}$, S.~Jordan$^{4}$, K.A.~Lawrie$^{1}$,
  A.E.~Lynas-Gray$^{11}$, T.~Maccarone$^{12, 13}$, T.~Marsh$^{2}$,
  S.~Parsons$^{14,2}$, R.~Silvotti$^{15}$, J.~Subasavage$^{16}$, S. Torres$^{6}$, P.~Wheatley$^{2}$}

\vspace{1cm}
\today

\end{center}
{\small 
*Contact email: slc25@le.ac.uk\\
$^{1}$Department of Physics and Astronomy, University of Leicester, University Road, Leicester, LE1 7RH,
  UK\\
$^{2}$Department of Physics, University of Warwick, Coventry CV4 7AL, UK \\
$^{3}$Institut de Ciències de l'Espai (ICE-CSIC/IEEC), Campus UAB, 08193 Bellaterra, Barcelona, Spain\\
$^{4}$Astronomisches Rechen-Institut, Zentrum für Astronomie der Universität Heidelberg, Mönchhofstr. 12-14, 69120, Heidelberg, Germany\\
$^{5}$Astrophysics Research Institute, Liverpool John Moores University, IC2, Liverpool Science Park, 146 Brownlow Hill, Liverpool L3 5RF, UK\\
$^{6}$Departament de Física Aplicada, Universitat Politècnica de Catalunya, c/Esteve Terrades, 5, 08860 Castelldefels, Spain\\
$^{7}$Scottish Universities Physics Alliance (SUPA), Institute for Astronomy, School of Physics, University of Edinburgh, Royal Observatory, Blackford Hill, Edinburgh EH9 3HJ, UK\\
$^{8}$Dr. Remeis-Sternwarte, Astronomisches Institut der Universität Erlangen-Nürnberg, Sternwartstr. 7, 96049 Bamberg, Germany\\
$^{9}$Lunar and Planetary Laboratory, Sonnett Space Sciences Bld., University of Arizona, Tucson, AZ 85721, USA\\
$^{10}$Armagh Observatory, College Hill, Armagh BT61 9DG, Northern Ireland\\
$^{11}$Department of Physics, University of Oxford, Oxford, UK\\
$^{12}$Department of Physics, Texas Tech University, Lubbock, TX 79409, USA\\
$^{13}$School of Physics and Astronomy, University of Southampton, Hampshire SO17 1BJ, UK\\
$^{14}$Departmento de Física y Astronomía, Universidad de Valparaíso, Avenida Gran Bretana 1111, Valparaíso 2360102, Chile\\
$^{15}$INAF-Osservatorio Astrofisico di Torino, Strada dell'Osservatorio 20, 10025, Pino Torinese, Italy\\
$^{16}$United States Naval Observatory - Flagstaff Station, 10391 W. Naval Observatory Rd., Flagstaff, AZ 86001-8521, USA\\

\normalsize

%% file: chapter_Introduction.tex
\setcounter{chapter}{0}
\bibliographyunit[\chapter]
\chapter[Introduction]{Introduction}
\section{General remarks}   

Gaia is a satellite mission of the ESA, aiming at absolute astrometric measurements of about one billion stars
($V<20$) with unprecedented accuracy \citep[see e.g.][and references therein]{Jordan2008} Additionally, magnitudes and colors will be obtained
for all these stars. Additionally,  near infrared (8470 -- 8740\,\AA) medium resolution spectra will be taken with a resolution of
$\mathrm{R} = \lambda / \Delta \lambda = 11\,500$, aiming at the determination of radial velocities for
for  bright objects ($V<17.5$). 

The orbit of the Gaia mission has been chosen to be a controlled Lissajous orbit around the Lagrangian point L2 of the Sun-Earth system
in order to have a quiet environment for the payload in terms of mechanical  and  thermo-mechanical stability. Another advantage of this position is
the possibility of uninterrupted observations, since the Earth, Moon and Sun all lay within Gaia's orbit.
The aim of Gaia is to perform absolute astrometry rather than differential measurements in a small field of view.
For this reason, Gaia -- like HIPPARCOS -- (i) simultaneously observes in two fields of view (FoVs) separated by a large
basic angle of 106.5$^\circ$, (ii) roughly scans along a great circle leading to strong mathematical closure conditions,
(iii) performing mainly one-dimensional measurements, and (iv) scanning the same area of sky many times during the mission
under varying orientations. 

These conditions are fulfilled by Gaia's nominal scanning law:
The satellite will spin around its axis with a constant rotational period of 6 hours. The spin axis will precess
around the solar direction with a fixed aspect angle of 45$^\circ$ in 63.12 days. On average, each object in the sky is transiting the
focal plane about 70 times  during the 5 year nominal mission duration. Most of the times, an object transiting through
one FoV is measured again after 106.5 or 253.5 minutes (according to the basic angle of 106.5$^\circ$) in the 
other FoV.

The Gaia payload consists of three instruments mounted on a single optical bench: The astrometric instrument,
the photometers, and a spectrograph to measure radial velocities.

The astrometric field consists of 62 CCDs and a star is measured on 8-9 CCDs during one transit.
The accumulated charges
of the CCD are transported across the CCD in time delay integration mode in synchrony with the images.
 In order to reduce the data rate and the read-out noise only small windows
around each target star, additionally binned in across-scan direction depending on the object's magnitude,
are read out and transmitted to the ground.

Multi-colour photometry is provided by 
two low-resolution fused-silica prisms dispersing all the
light entering the field of view in the along-scan direction prior to detection.
The Blue Photometer (BP) operates in the wavelength
range 3300--6800\,\AA; the Red Photometer (RP) covers the
wavelength range 6400--10500\,\AA.  

The RVS is a near infrared (8470 -- 8740\,\AA), medium resolution spectrograph:
$\mathrm{R} = \lambda / \Delta \lambda = 11\,500$. It is illuminated by the same
two telescopes as the astrometric and photometric instruments.  

The  astrometric core solution will be based on about  $10^8$ primary stars which means to solve for
some $5\times 10^8$ astrometric parameters (positions, proper motions, and parallaxes). However, the attitude of the
satellite (parameterized into $\sim 10^8$ attitude parameters over five years) can also only be determined with high 
accuracy from the measurements itself. Additionally, a few million calibrational parameters describe the geometry of the
instruments.

Gaia was launched in December 2013 The launch vehicle is a Soyuz-Fregat rocket which will lift-off from Sinnamary in French Guyana.
 After a several-month commissioning phase Gaia will start its five year period of
nominal measurements which can be extended by another year. About two to three years after the mission the final Gaia catalogue will be published.
However, it is foreseen that several intermediate catalogues will be produced well before. After about 18 months of the measurements, a first 
astrometric solution will be  possible which can solve  for positions, proper motions and parallaxes of almost all stars with reduced accuracy.

The accuracy of the astrometric measurements depend on the brightness and spectral type of the stars. At $G=15$ mag (Gaia magnitude, approximately equivalent
to the $V$ magnitude) the final accuracy in position, proper motion per year and the parallax will be $25$ micro arcsecond. A somewhat larger accuracy is
reached for red stars. At $G=20$ mag the final precision drops to about $300\,\mu$as. 

The end-of-mission photometric performance is of the order of 5-15 mmag, again depending on the brightness and spectral type.

\section{Gaia's performance for white dwarfs}   
\cite{Torresetal05} have performed intensive Monte-Carlo simulations and arrived at about 400,000 white dwarfs
down to $G\approx V=20$ that will be detected by Gaia. For disk white dwarfs Gaia will be practically complete up to
100\,pc and will observe about half  of all white dwarfs within  300\,pc, decreasing to one third at distances of
400\,pc.  Disk white dwarfs at the cut-off of the luminosity function ($M_{\rm bol}\approx 15.3$, $M_V\approx 16$) can be
detected up to distances of 100\,pc, considerably improving the age determination of the solar neighborhood to about
$\pm 0.3$\,Gyr. Moreover, a detailed check of white dwarf cooling theory is possible by a careful analysis of the
Gaia white dwarf luminosity function.

Additionally to the astrometric information mentioned above, the analysis of the BP/RP low-resolution spectra will be essential.
If we add up all BP spectra for a give white dwarf for the entire mission, Balmer lines can be detected down to about $G=18$
in hydrogen-rich (DA) white dwarfs close to the maximum of the Balmer line strength ($T_{\rm eff}\approx 12,000$\,K. For other
atmospheric parameters the shape of the BP and RP spectra can be used to determine approximate effective temperatures. This 
information can be used to select the white dwarfs from the whole sample of observed stars.

However, a detailed analysis of white dwarfs will need follow-up spectroscopy  in order to
determine their precise atmospheric parameters. Only with this information the scientific programmes described in this
white book can fully exploit the valuable data from the Gaia satellite.

One  of the major advantages of the Gaia sample of  white dwarfs
is that it constitutes  an all-sky survey with very clear selection criteria. Biases introduced by 
unclear selection effects are certainly one of the major obstacles of statistical investigations of white
dwarfs \citep{Jordan2007}. One can clearly conclude that white dwarf research will tremendously benefit from the Gaia data.

This white paper will provide an overview on the scientific exploitation of the Gaia data of white dwarfs. We will show which broad range of 
topics is connected to this research, from exoplanet research to a better understanding of our galaxy, from the improvement of our knowledge of the
late phases of stellar evolution to test of physical theories.

\putbib[introduction_refs]
\bibliographyunit[\section]

%% file: chapter_Science.tex
\setcounter{chapter}{1}
\chapter[Science]{Science}

\input{section_MR_relation}
\input{section_IFMR}
\input{section_LF}
\input{section_mag_fields}
\input{section_pulsations}
\input{section_binaries}
\input{section_planets}
\input{section_late_stages}
\input{section_projects}

%% file: section_MR_relation.tex
\section{Mass-Radius relation}
Two of the most important physical parameters that can be measured for any star are the mass (M)
 and radius (R). They determine the surface gravity (g) by the relation  given in equation \ref{MR1} 
\begin{equation}
\label{MR1}
g=\frac{GM}{R^2},
\end{equation}

where  G is the gravitational constant.
 Hence, if log g is measured the mass can be calculated provided the stellar 
radius is known. One outcome of Chandrasekhar's original work on the structure
 of white dwarfs was the relationship between mass and radius, arising from the
 physical properties of degenerate matter. Further theoretical work yielded the
 Hamada-Salpeter zero-temperature mass-radius relation \citep{hamada61}.
 However, white dwarfs do not have zero temperature, indeed many are very hot. 
Hence, the Hamada-Salpeter relation is only a limiting case and the effects of 
finite temperature need to be taken into account. Several authors have carried 
out evolutionary calculations, where the radius of a white dwarf of given mass 
decreases as the star cools. Those of \citet{wood90, wood95}, have a semi arbitrary 
starting point for the hottest models while the \citet{blocker97} models are
 full evolutionary calculations from the AGB. Other calculations are available 
from \citet{althaus97,althaus98} and \citet*{fontaine01}.
	
\begin{figure}[ht]
\begin{minipage}[b]{0.25\linewidth}
\centering
\includegraphics[height=.4\textheight, angle=0]{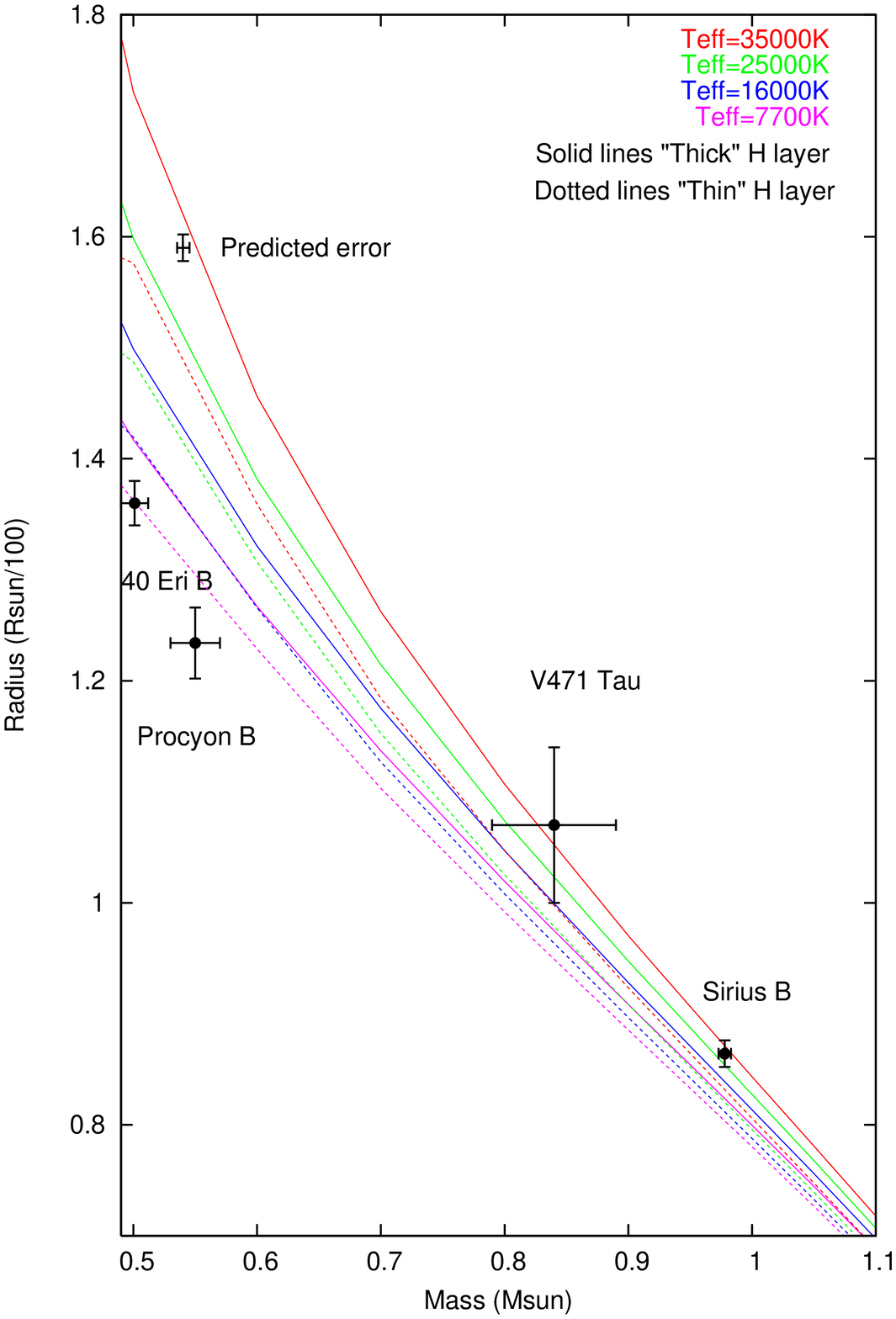}
\end{minipage}
\hspace{3.0cm}
\begin{minipage}[b]{0.25\linewidth}
\centering
\includegraphics[height=.4\textheight, angle=0]{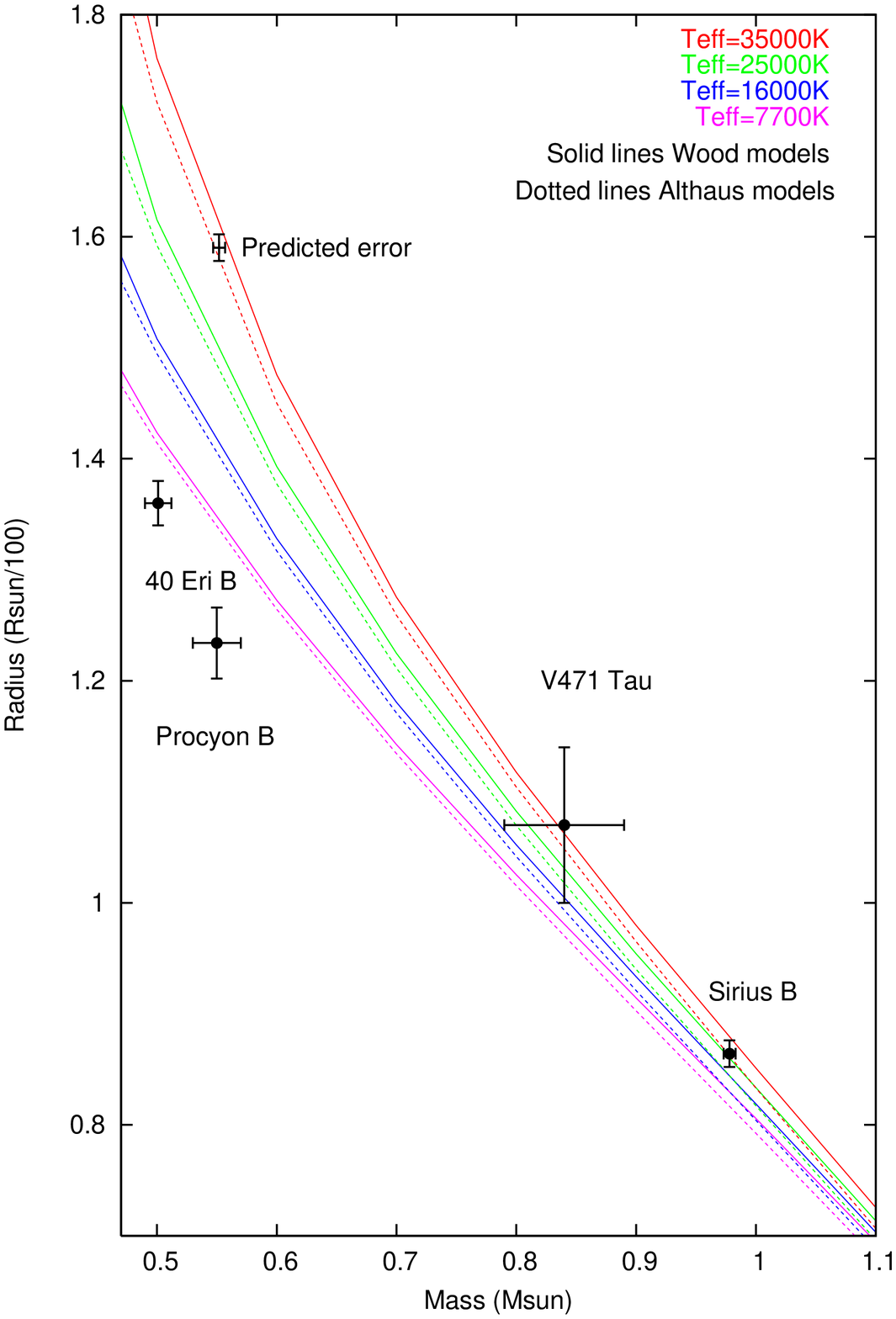}
\end{minipage}
\caption[Mass-Radius data, and model predictions]{\label{MRfig1}Left:\citet{fontaine01} models ``thick'' (solid) and ``thin'' (dashed) hydrogen envelopes. Right:Models from \citet[solid]{wood90} and \citet[dashed]{althaus98}.The data for 40 
Eri B, Procyon B, V471 Tau and Sirius B are respectively from \citet{provencal98,provencal02},
 Bond et al (2011; private comm), \citet{girard00}, \citet{obrien01} and \citet{barstow05}.}
\end{figure}

For the largest group of white dwarfs, the H-rich DA stars, there is a very powerful 
technique to determine log g and effective temperature (T$_{\rm eff}$) based on fitting 
synthetic stellar spectra to 
the hydrogen Balmer absorption line profiles. However, the resulting estimates of 
the mass and radius have generally relied on using the theoretical mass-radius 
relation to remove the inter-dependence of M and R on g. While, the basic 
stellar models and white dwarf mass-radius relation are not in serious doubt,
 there are uncertainties that arise from the detailed input physics and
 higher-level refinements that take into account the finite stellar 
temperature and details of the core/envelope structure. Varying the 
assumed input parameters in these models can lead to quite subtle, but 
important differences in the model predictions as illustrated in Figure \ref{MRfig1} which 
shows several different theoretical mass-radius relations along with the observed masses 
and radii of the white dwarfs in four well studied astrometric binaries.

\subsection{Testing the mass-radius relation and stellar models}

Direct observational tests are difficult. To break the reliance of mass and radius determinations on 
the theoretical models requires additional independent information on at least one of M or R. For 
example, accurate photometry of the stellar flux (F$_{\lambda}$ allows determination of R, provided we know the distance (D) as shown in equation \ref{MR2}

\begin{equation}
\label{MR2}
F_{\lambda} = 4\pi\left(\frac{R^2}{D^2}\right)H_{\lambda},
\end{equation}
where H is the Eddington flux.
Alternatively, if we can determine the gravitational redshift (V$_{gr}$) of a white dwarf we get a
 second relation between M and R, allowing us to solve for their values as shown in equation \ref{MR3}
\begin{equation}
\label{MR3}
V_{gr}= 0.636\frac{M}{R},
\end{equation}
where M and R are in Solar units and V$_{\rm gr}$ is in kms$^{-1}$.

However, V$_{\rm gr}$ can only usually be measured independently if the white dwarf is in a binary system as its space velocity needs to be known.

\begin{figure*}
\begin{center}
\includegraphics[height=.5\textheight, angle=90]{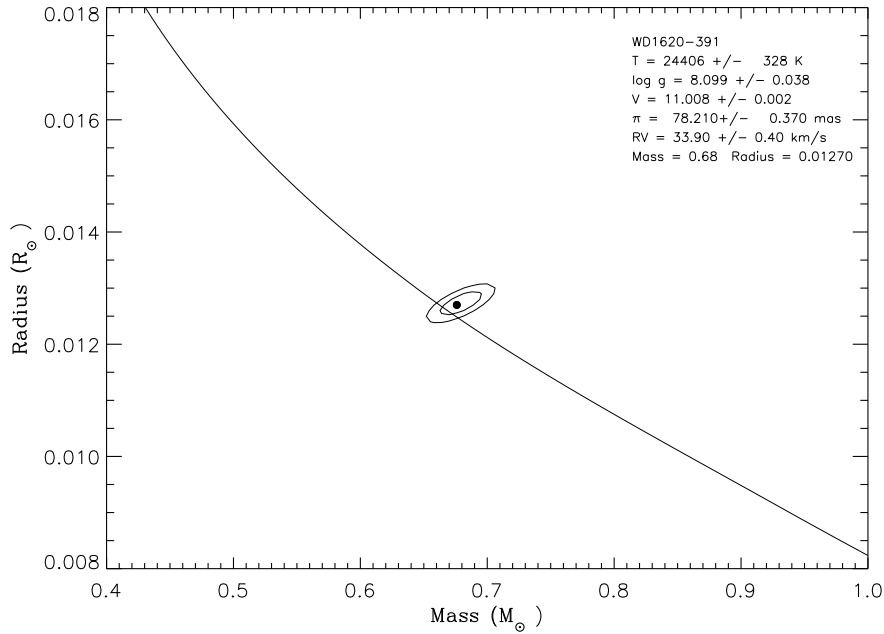}
\caption{\label{MRfig2}DA2.1 + G5V system WD1620-391 (CD-38$^{\circ}$ 10980) and a theoretical mass-radius relation. The one and two sigma contours sit directly astride the theoretical mass-radius relation for a 24,400 K white dwarf.}
\end{center}
\end{figure*}

Figure \ref{MRfig1} represents the recent state of the art for independent determinations of M and R which can be compared with a range of temperature dependent theoretical mass-radius determinations.  It includes a relatively  small number of white dwarfs in well known binary systems.   Recently \citet{holberg12} have presented similar results for a dozen systems using recent parallaxes, improved photometry and mass-radius relations specific to the temperature of each star.   The method used was to express the stellar radius in the three equations, \ref{MR1}, \ref{MR2}, \ref{MR3}, as explicit functions of mass and use each to independently to define a three-term chi-square function which can be minimized with respect to the 'observed' radii and their corresponding uncertainties.  The one and 2 sigma uncertainty contours can then be plotted in the mass-radius plane and directly compared to the theoretical mass-radius relation for each star.    A particularly good example is shown in Figure \ref{MRfig2} for  well known DA2.1 + G5V system WD1620-391 (CD-38$^{\circ}$ 10980), where the one and two sigma contours sit directly astride the theoretical mass-radius relation for a 24,400 K white dwarf.  The important point of this demonstration is that the mass and radius were determined with no direct dependence on theoretical mass-radius relations. 
Several important conclusions follow from \citet{holberg12}.   First is that for those stars where meaningful comparisons were possible the 'thick' hydrogen layer mass-radius relations match the data reasonable well.  The only exception was 40 Eri B (WD 0413-077), where the mass and radius clearly fit a 'thin' hydrogen layer mass but not a 'thick' hydrogen layer.  Secondly, for most stars,  current Hipparcos and ground based parallaxes are the limiting factors in such analyses.   Indeed, if relative parallax errors can be reduced below the 1\% level other factors such as photometric, gravitational redshift and surface gravity uncertainties dominate and results similar to that shown in Figure \ref{MRfig2} can be achieved for a great many stars.  This is where Gaia can play a crucial role.   In principle it will be possible to observationally define a precise empirical mass-radius relation for as many as perhaps 100 white dwarfs which can then be compared with theoretical mass-radius relations for stars of different core compositions and hydrogen layer masses.   In anticipation of Gaia parallaxes a program to obtain accurate gravitational redshifts for a large sample of suitable stars will soon be underway.   Many of these systems will automatically become prime candidates for the determination of the initial mass-final mass relation (IFMR, see Section 2.2). 

\subsection{Calibration and the Mass-Radius Relation}

White dwarf stars and in particular pure hydrogen DA white dwarfs have long been used as ground-based and space-based photometric calibrators.  With Gaia parallaxes it will be possible  to do two things.  First, as shown by \citet*{Holberg06} and \citet*{Holberg08} it is presently possible, using existing parallaxes, to establish a one-to-one correlation between observed absolute magnitudes and synthetic absolute magnitudes for DA white dwarfs.  The synthetic magnitudes are calculated from the spectroscopic effective temperatures and gravities and need only be normalised at a single observed magnitude, say $V$-band.   The photometric scale is that presently used is that established for the Hubble Space Telescope, which has an absolute accuracy of 1 \% at visual wavelengths.  The existing correlation between observed fluxes and synthetic fluxes in well characterised bands is also at the 1 \% level (see Figs. 1 and 2 of \citealt{Holberg08}).  At present the largest source of uncertainty in these correlations are the existing parallaxes.  Gaia parallaxes will virtually eliminate this source of uncertainty allowing the enhanced correlations to be extended to thousands of stars.  In  way this way accurate absolute fluxes can be established for a large ensemble of standard stars that can be reliably  used for space-based and ground-based calibrations.

The empirical relations between observed and synthetic absolute fluxes described above leads to the second way that Gaia parallaxes can be used investigate fundamental stellar astrophysics; observationally testing the degenerate mass-radius  relation.   Basically, Gaia parallaxes can be used to produce accurate estimates of white dwarf radii through the stellar solid angle and Eddington  fluxes.    Using such radii in binary systems containing white dwarfs, where the gravitational redshift can accurately be determined, represents  a powerful opportunity to critically test basic temperature dependent mass-radius relations.    Unless eclipsing systems are available, white dwarf radii are most easily estimated from absolute photometry and accurate parallaxes.  If such radii are coupled with independent constraints on M and R from spectroscopic determinations of temperature and gravity and  gravitational redshifts, then three independent measures of M and R are possible.   Using these constraints and the associated uncertainties in M and R it is possible to construct contours of equal likelihood in the mass-radius plane; see Fig. 4 of \citet{Farihi11} for an example of this type of analysis.  

The particular programme advocated here is to prepare for the arrival of Gaia parallaxes by identifying a large sample of up to 100 binary systems containing DA white dwarfs where accurate photometric fluxes and spectroscopic, temperatures, gravities and redshifts can be obtained (a few suitable systems and observations already exist).  These DA stars would be selected to densely and uniformly sample a range of masses between 0.5 Solar masses up to near the Chandrasekhar limit.  The goal would be to systematically obtain the necessary ground-based observations, so that when Gaia parallaxes are available, independent mass and radius estimates can be promptly compared with theoretical  mass-radius relations at the few percent level.   This should allow robust statistical tests at the one percent level.

\subsection{Eclipsing white dwarf / main-sequence binaries and the M-R relation}

\begin{figure}
\includegraphics[width=0.35\textwidth]{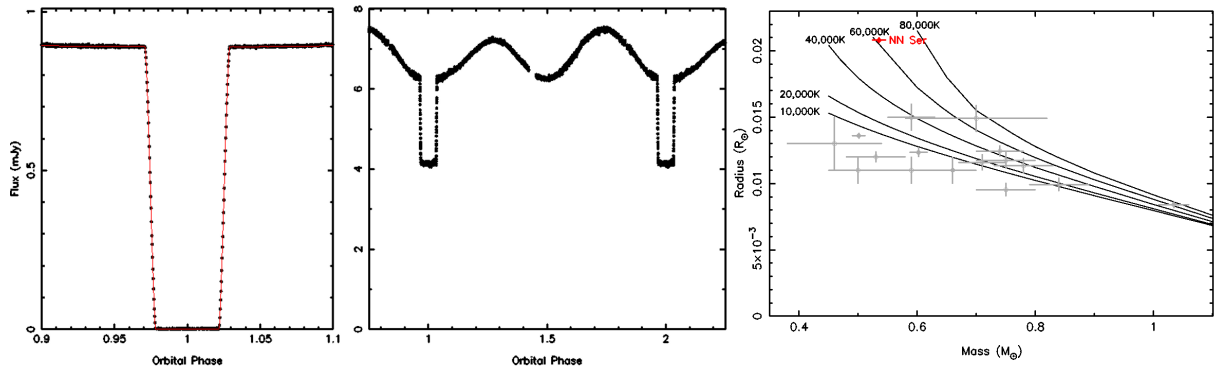}
\caption[PCEB light curve and fits]{{\it Left:} Primary eclipse of the PCEB NN~Ser from \citet{Parsons10a} 
with model fit over-plotted. The total eclipse duration is 12 minutes. In this 
case the radii of the two stars were measured to a precision of $1\%$. {\it Centre
:} Light curve of the PCEB QS~Vir from \citet{Parsons10b} showing out-of-eclipse 
variations due to the tidal distortion of the main-sequence star by the white 
dwarf. These effects can be used to determine the period of the system even in 
the absence of an eclipse. {\it Right:} Mass-radius relation for white dwarfs. 
The red point represents the mass and radius of the white dwarf in the PCEB NN~Ser
. It is the most precise model-independent measurement of a white dwarfs mass and
radius to date.}
\label{pceb_fig}
\end{figure}

White dwarf plus main-sequence binaries are numerous throughout our Galaxy. Post
common envelope binaries (PCEBs) are a type of close white dwarf / main-sequence
binary with periods $\leq 1$ day. A small number of these systems are inclined in
such a way that, as viewed from Earth, they exhibit deep eclipses, as the main 
sequence secondary star passes in front of the white dwarf. These
deep eclipses allow us to measure extremely precise radii which, combined with 
spectroscopic observations that yield the component masses, enable us to test 
mass-radius relations for both the white dwarf and its, often low-mass, main-
sequence companion. A typical eclipse of a PCEB is shown in the left hand panel 
of Figure~\ref{pceb_fig}. The sharp ingress and egress features allow us to measure
radii to a precision of $\sim1\%$ in a way that is almost entirely independent of 
model atmosphere calculations.

We currently know of over 2000 white dwarf / main-sequence binaries 
\citep{Rebassa11}. Of these there are 34 confirmed eclipsing systems. Gaia is 
expected to discover around 1000 more, giving us a large population with well 
understood selection effects. It will also provide orbital periods; this will be
achieved by detecting multiple eclipses, but also by providing high S/N photometry
out of eclipse. Eclipsing PCEBs often show large ($\sim 0.1-0.3$ mag) variations out 
of eclipse due to irradiation or tidal distortion of the main-sequence star as shown
in the centre panel of Figure~\ref{pceb_fig}. These variations can even be used to 
determine the orbital periods of non-eclipsing systems. Gaia will also provide radial
velocity data on the main sequence star for the brighter systems (around the calcium infrared triplet). Spectral
features from the main-sequence companion are likely to be visible in this window
providing some constraints to the masses of the two stars.

Additional data are required to determine precise masses and radii. No spectral 
features from the white dwarf will be visible in the Gaia radial velocity spectrum, 
therefore, in order to
measure the radial velocity of the white dwarf (and thus the masses) follow-up 
phase-resolved spectroscopy around the hydrogen Balmer lines will be needed. 
For the most part, Gaia data will not be adequate for precision analysis of eclipses
and follow-up high-speed photometery will be required to measure precise radii.
This combination of spectroscopic and photometric data can lead to masses and radii
of the order of precision of those shown in the right hand panel of 
Figure~\ref{pceb_fig}, good enough to test white dwarf mass-radius relations. These
eclipsing systems can also be used to test mass-radius relations of low-mass stars
down to brown dwarf masses.

\putbib[MR_refs]

%% file: section_IFMR.tex
\section{Initial mass-final mass relation (IFMR)}

The initial mass-final mass relation (IFMR) is a theoretically predicted positive correlation between the main sequence mass of a star
with M$\simless$10 $M_{\odot}$ and the mass of the white dwarf remnant left behind after it has expired (e.g. \citealt*{iben83}). 
This relation is of paramount importance for several fields in 
astrophysics (e.g. ages and distances of globular clusters, chemical 
evolution of galaxies, white dwarf population...) and understanding its form provides a handle on the total amount of gas enriched with He, N and other
metals that  95$\%$ of all stars, return to the interstellar medium at the end of their lifecycles (e.g. \citealt*{carigi99}).
Moreover, the form of the upper end of the IFMR is relevant to studies of Type II supernovae as it can provide a constraint on the 
minimum mass of star that will experience this fate. For example, with robust constraints on this mass, the observed diffuse neutrino background can serve
better as an empirical normalisation check on estimates of the star formation history of the universe (e.g. \citealt*{hopkins06}).

\begin{figure}
\begin{center}
\includegraphics[scale=0.5,angle=270]{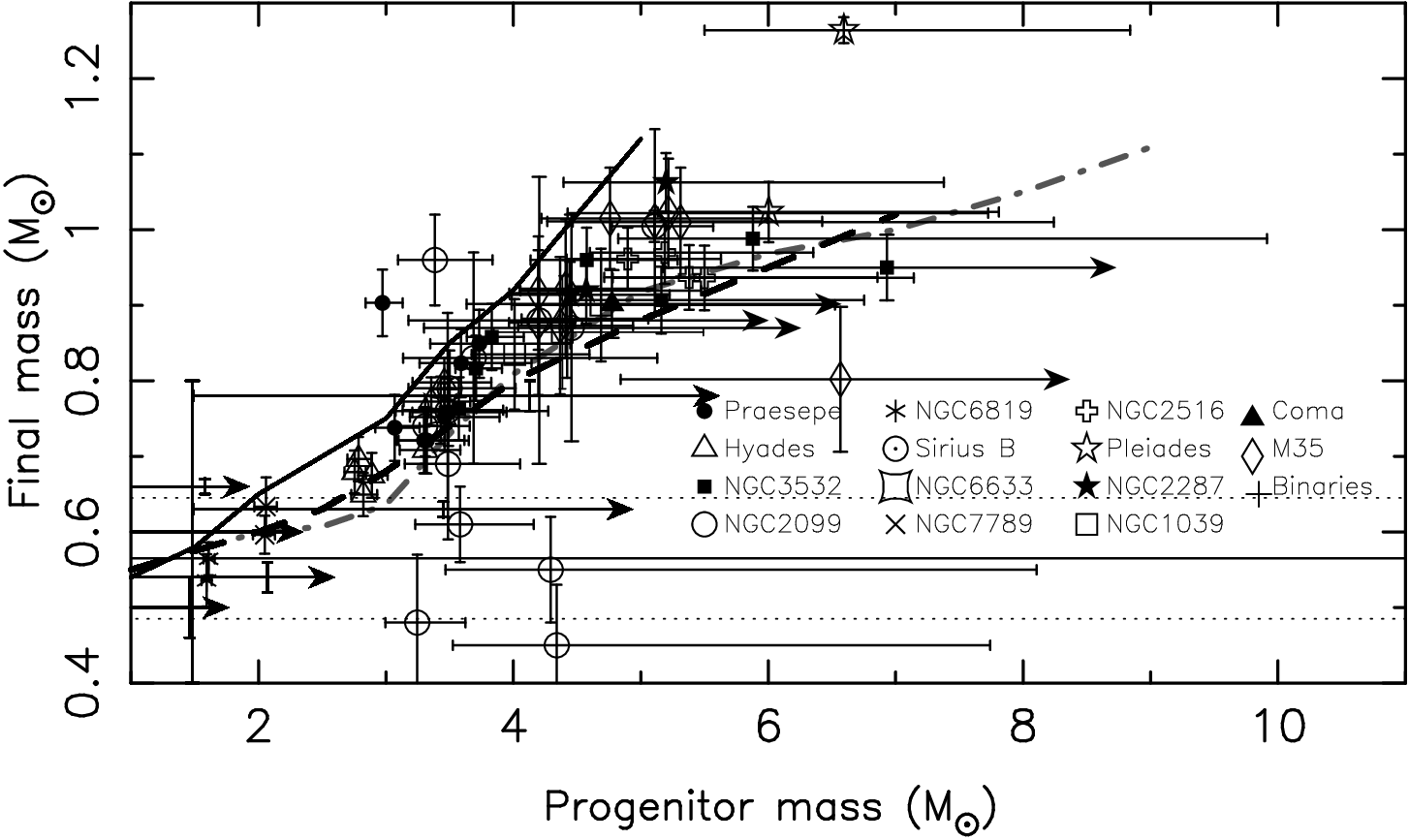}
\caption[Semi-empirical IFMR]{Final masses versus initial masses of the available cluster and 
wide binaries data.The dashed black line is the semi-empirical \citet{weidemann00} IFMR, the thick solid line is the IFMR as given by the \citet{girardi00} models and the grey dot-dashed line is the initial mass-core mass at the first thermal pulse relation from \citet*{karakas02}. The peak in the field white dwarf mass distribution (thin solid line) and $\pm$1$\sigma$ is represented by the thin dotted lines.  }
\label{fig}
\end{center}
\end{figure}

The form of the IFMR is extremely difficult to predict from theory alone due to the many complex processes occurring during the final phases
of stellar evolution (e.g.\citealt*{iben83}). There have been several attempts to produce a theoretical IFMR (\citealt{dominguez99,marigo01,marigo07}, but
differences in evolutionary codes can lead to very different results (see \citealt{weidemann00} for a review). 
The IFMR is generally supposed to be a linear relation over the range 1.16M$_{\odot}<$M$_{\rm init}<6.5$$M_{\odot}$
\citep{kalirai08,catalan08b,casewell09,dobbie09}, but there is some evidence that the IFMR is steeper between 3 $M_{\odot}$  and 4 $M_{\odot}$ than for initial
masses either side of this value (see Figure \ref{fig}; \citealt{dobbie09}). There is a problem with fitting a linear IFMR to the current data, and that is that
the fit  is affected by the high mass white dwarfs which tend to have large error bars on their mass determinations (driven by the steep change in the
relationship between stellar mass and lifetime at these young ages; \citealt*{williams09}) and the dearth of data in the low mass regime
\citep{catalan08,dobbie09,salaris09}.

The first attempt to empirically map the IFMR was made by \citet{weidemann77} who obtained the masses of the white dwarfs in the Hyades and Pleiades from fitting synthetic profiles to the Balmer lines. Open cluster white dwarfs were used because 
the total age of a white dwarf can be expressed as the sum of its 
cooling time and the main-sequence lifetime of its progenitor. This  
latter parameter depends on the metallicity of the progenitor of the white dwarf,information 
which is lost once the star becomes a white dwarf. Since this method combines observational data and the use 
of models, the obtained relationship is, in effect semi-empirical. Using cluster white dwarfs is the most traditional way to define the IFMR (e.g.Figure \ref{fig};\citealt{weidemann87,weidemann00, ferrario05, dobbie06, williams07, kalirai08,rubin08, casewell09,dobbie09}), nonetheless, until relatively recently, rather few white dwarf members of open star clusters had been identified: 61 white dwarfs from 12 open star clusters \citep{williams09}.  The uncertainties in membership status, cluster ages and the relatively large distances involved resulted in large scatter in 
the IFMR \citep{claver01, ferrario05}. Another problem with this method is that nearby open star clusters have to be sufficiently rich and old enough to harbour detectable white dwarfs. This means that the semi-empirical IFMR is relatively well populated only between 2.5 and 7.0M$_{\odot}$.

An alternative way to expand the parameter space of the IFMR is to use white dwarfs in binary systems where it can be assumed that 
 the members of a wide binary (common proper 
motion pair) were born simultaneously and with the same chemical 
composition (\citealt*{wegner73,oswalt88}). Since the components are 
well separated (100 to 1000 AU), it can be considered that they have evolved as isolated stars.  \citet{catalan08} used these pairs to provide additional 
points on the IFMR by obtaining the total age of the white dwarf and the metallicity of its progenitor
 from the study of the companion star. However, only a small 
sample of binaries were used to improve the initial-final mass 
relationship, because a parallax measurement is necessary in order to 
obtain the luminosity of the companion with accuracy, and from this the 
total age of the system.These systems are in general, much closer than open star clusters, which means they are brighter
and thus much easier to study spectroscopically. Of these 6 additional white dwarfs (+ in Figure \ref{fig}), 3 have masses of less that 2M$_{\odot}$. There are however, large errors on some of these measurements, and they also show the dispersion that appears in the cluster based IFMR. The reason for this dispersion in both cases is currently unknown (i.e. it is not metallicity based).
One other option for the IFMR is to use double degenerate systems such as PG0922+16 \citep*{finley97}, where the difference in cooling times between the two white dwarfs can be used to calculate the stellar lifetimes and hence initial mass. This method is however, limited by the relative scarcity of double degenerate systems.

To better define the IFMR as it is now, we require two things; firstly more information about the white dwarfs we have; e.g. 
parallax to confirm cluster membership and distances to binaries,  and secondly, smaller errors on the ages of these objects. \citet{salaris09} 
observed that the largest source of systematic error in IFMR calculations, is from the uncertainty in the cluster age used to define it.
 Gaia spectroscopy and parallax will allow much better determination of white dwarf parameters, particularly in southern open clusters 
that have not been observed by large scale optical photometric surveys such as the SDSS.

To expand the IFMR at both the low and high mass ends, we require more white dwarfs in known age systems. 
\citet{torres05} used Monte Carlo simulations to predict $\sim$ 400000 white 
dwarfs will be detected by Gaia. Of these, 25$\%$ are expected to be 
in binary systems,  6.5$\%$ belonging to wide binaries with a K-type 
or earlier companion \citep{holberg08}, though the statistics are not 
complete.  Gaia will provide accurate parallaxes for all these systems 
which will be essential to obtain their luminosity with precision.  
If the predictions are correct Gaia will provide thousands of 
wide binaries useful to improve the semi-empirical initial-final mass 
relationship. In particular, this will benefit the coverage of the 
low-mass domain (Figure \ref{fig}) which represents 90$\%$ of the stellar population (white dwarfs with masses $\sim$ 
$0.6M_{\odot}$).

 Gaia will provide a large range of both parallax and proper motions for the whole sky, 
allowing us to expand our parameter space by using both clusters and binaries. We will also be able to use the
 proper motions and parallax with the convergent point method  to identify 
white dwarfs as part of moving groups, and hence obtain an age for them as for GD50 \citep{dobbie06b}. 

For every new white dwarf discovered by Gaia we will have a proper motion, photometry and a parallax. Gaia will also provide radial velocities of the primary star for binary systems, and cluster members. We will also require high signal to noise ratio spectroscopy to obtain an effective temperature, gravity and mass, estimate of whether the white dwarf is magnetic, and  radial velocity which will confirm association membership (of a cluster or binary). These data will allow us to reject objects that are not true cluster or binary members, resulting in less scatter in the IFMR.

Spectra of the cluster members and primary stars will also be required to give an estimate of metallicity, which along with magnetic objects may cause uncertainties in the IFMR.  These data will come from multi-object spectrographs such as UVES$+$GIRAFFE as part of the ESO Gaia survey, or the spectrographs BigBOSS being proposed for the  Kitt Peak 4m, HERMES being proposed for the AAT and WEAVE for the WHT. These instruments will provide
accurate chemical abundance measurements for open star clusters, leading to a much cleaner main sequence, due to rejection of outliers/non-members, and hence a more accurate age estimate for the cluster.

\putbib[IFMR_refs]

%% file: section_LF.tex
\section{Luminosity function}

%%%%%%%%%%%%%%%%%  Useful definitions

\def\pg{(p,\gamma)}
\def\pa{(p,\alpha)}
\def\gp{(\gamma,p)}
\def\ap{(\alpha,p)}
\def\ag{(\alpha,\gamma)}
\def\ga{(\gamma,\alpha)}
\def\msun{M_{\odot}}
\def\lsun{L_{\odot}}
\def\rsun{R_{\odot}}
\def\rns{R_{NS}}
\def\msuny{\,M_{\odot}{\cdot}yr^{-1}}
\def\gcm{\,g{\cdot}cm^{-3}}
\def\gcmc{\,g{\cdot}cm^{-2}}
\def\gcms{\,g{\cdot}cm^{-2}{\cdot}s^{-1}}
\def\km{\,km}
\def\kms{\,km{\cdot}s^{-1}}
\def\gk{\,GK}
\def\ergs{erg \cdot s^{-1}}
\def\sec{\,s}
\def\macc{M_{acc}}
\def\kev{\,keV}
\def\mev{\,MeV}

\def\sun{\odot}

%%%%%%%%%%%%%%%%%  Journals

\def\aj{AJ}
\def\apj{ApJ}
\def\aap{A\&A}
\def\apjl{ApJ}
\def\apjs{ApJS}
\def\aaps{A\&AS}
\def\nat{Nature}
\def\araa{ARA\&A}
\def\apss{Ap\&SS}
\def\mnras{MNRAS}
\def\aapr{A\&A~Rev.}
\def\ao{Appl. Optics}
\def\prd{Phys. Rev. D}
\def\prl{Phys. Rev. Lett.}
\def\ssr{Space Sci. Rev.}
\def\jcps{Jour. of Comp. Phys.}
\def\cqg{Class. \& Quantum Grav.}
\def\pasp{Pub. Ast. Soc. Pacific}
\def\jpcs{Jour. of Phys. Conf. Ser.}
\def\sun{\odot}

\subsection{Age of the Galaxy} 

The field of white dwarf (WD) cosmochronometry~\citep[e.g.][]{1998ApJ...497..870W} is an increasingly active area of
research in Galactic stellar astrophysics. Put simply, there exists a minimum absolute
luminosity for the WD remnants of a given stellar population of finite age. A plot of
the number of WDs per unit volume per absolute magnitude interval (the WD luminosity
function) will show, in principle, a sharp cut--off at the faint end below which WDs
are no longer observed~\citep*{1988ApJ...332..891L}. 

It is only comparatively recently that digitised sky surveys have reached the depth
and time coverage required to extract large samples of cool WDs via the technique
of reduced proper motion~\citep[e.g.][]{2006AJ....131..571H}. 
Recently,~\citet{2011arXiv1102.3193R} -- hereafter RH11 --
report the largest sample achieved to date based on the legacy
Schmidt all--sky photographic surveys. Despite the limited passband coverage
(photographic $B_JR_FI_N$) and image quality (seeing typically $>2$~arcsec),
a sample of $\sim10,000$ WDs was assembled covering $3\pi$~str of sky with accurately
characterised completeness and reliability. 

Despite recent advances, however, there remain several important
but unresolved issues regarding the Galactic WDLFs and associated age determinations. 
Firstly, the ages are only poorly constrained because the volume sampled for the 
faintest and coolest (and
therefore oldest) WDs is small due to magnitude limited surveys (R$_F$=19--20).
Secondly, limited passband coverage allows photometric distance determinations accurate only 
to $\sim50$\%. This smears out features in the LFs, and forces us to assume
atmospheric compositions for the candidate WDs, preventing accurate accounting of 
the effects of H or He--dominated atmospheres; coupled with the previous problem,
these result in uncertainties in the interpretation of the limited observational data at the faint
end of the WDLF~\citep[e.g.][]{2008ApJ...678.1279B}.

The combination of Gaia trigonometric parallaxes with next--generation ground--based surveys
will make a huge impact on studies of the faint end of the WDLF. While the sample size of
cool WDs will not increase dramatically from Gaia data alone (because current surveys are
magnitude--limited at $m\sim20$ as opposed to proper motion limited), new trigonometric
parallax relationships for WDs will be produced from large samples of all atmosphere types,
and those relations can be applied to much larger samples from deeper ground--based surveys.

RH11 present a new technique
that enables a decomposition of the major kinematic population WDLFs
(thin disk, thick disk and spheroid). This not only enables considerably more
reliable disk age estimates by removal of contaminating WDs from older populations
(a problem pointed out by several researchers, e.g.~\citealt{2005ARAandA..43..247R}, but which remained
unaddressed until this study), but of course also enables age determinations for those
older populations for the first time. For example, 
a WDLF study using data from Pan-STARRS~\citep[e.g.][]{2010PASP..122.1389B} and 
Gaia~\citep[e.g.][]{2007ASPC..372..139J} will advance 
WDLF cosmochronometry to `precision' levels for all three 
major kinematic components of the Galaxy. Using simple scaling
arguments, we can calculate by what factor we can increase WDLF samples over those
in RH11. Volume sampled for a uniformly distributed population (e.g.~the local
spheroid population) will go as $d^3$ for distance $d$ limited by magnitude
($d_{\rm PS}/d_{\rm RH11}=10^{(r_{\rm PS}-r_{\rm RH11})/5}$) and proper motion
($d_{\rm PS}/d_{\rm RH11}=\mu_{\rm RH11}/\mu_{\rm PS}$) where $r$ is the 
magnitude limit for the PanSTARRS (PS) and RH11 surveys, and $\mu$ is similarly
the limiting proper motion. Current measurements from survey data indicate
$r_{\rm PS}=r_{\rm RH11}+1.5$, at which the proper motion precision is likely
to be comparable with that of the (50\% incomplete) RH11 survey at it's detection limit.
Hence the number of PS spheroid WDs will be $\sim2\times2^3=16$ times
higher than in RH11. For the thin disk component, scale height effects reduce the
distance exponent to~2, so the factor increase is $2\times2^2=8$; the thick
disk increase will be somewhere  between the two. Assuming the analysis of~\citet{1998ApJ...497..870W},
the ages of all three components will be 
determined to a few percent in terms of the statistical uncertainty. Contrast
this with the present situation, where the thin disk is known to $\sim10$\%
at best~\citep{2010ApJS..190...77K} but with a larger systematic uncertainty due to contaminating
WDs from the older populations, while the thick disk and spheroid ages are
as yet undetermined.

Such a study will do much more than measuring the ages of the three
kinematic components. The availability of $grizy$ photometry from PS,
in conjunction with Gaia trigonometric parallaxes and
UKIDSS/VISTA infrared data for large subsamples, will enable
a finer analysis of atmospheric compositional effects leading to better photometric
distance estimates for the full magnitude limited sample. 
Hence, starburst features in all WDLFs, if present, will be
more pronounced leading to a better understanding of the star formation histories
of the progenitor populations. The elusive, but fascinating `ultra cool' WDs
exhibiting unusual SEDs~\citep[][~and references therein]{2008MNRAS.385L..23R} 
will be prevalent in the new sample, 
leading to a much better understanding of their nature and effect on the faint end of the
WDLF via follow--up studies. Finally, it will be possible to make an observational
census (as opposed to a theoretical model extrapolation) of the total number of
cool, very low luminosity stellar remnants in the disks and spheroid and a
measurement of their contribution to the total baryonic mass of the Galaxy, with
implications for interpretation of microlensing experimental results that invoke
$\sim0.5 {\rm M}_{\odot}$ stellar remnants as the lensing 
candidate~\citep{2005AandA...443..911C}.

%and {Paulin-Henriksson}, S. and {An}, J. and 
%    {Baillon}, P. and {Belokurov}, V. and {Carr}, B.~J. and {Cr{\'e}z{\'e}}, M. and 
%    {Evans}, N.~W. and {Giraud-H{\'e}raud}, Y. and {Gould}, A. and 
%    {Hewett}, P. and {Jetzer}, P. and {Kaplan}, J. and {Kerins}, E. and 
%    {Smartt}, S.~J. and {Stalin}, C.~S. and {Tsapras}, Y. and {Weston}, M.~J.

\subsubsection{Age of the Galactic disk}

A recent  assessment of the total  number counts of  disk white dwarfs
using Monte  Carlo techniques was done by  \cite{torr05}.  Using their
data  we pay  attention to  some specific  matters regarding  the disk
white  dwarf  luminosity  function.   In  particular we  ask  to  what
precision the age  of the Galactic disk can  be estimated.  They found
that the sample of disk  white dwarfs that Gaia will eventually detect
is almost  complete in magnitude  up to $G  \simeq 20$, and  all white
dwarfs within  this sample will have measurable  proper motions. Using
the $1/V_{\rm max}$ method  \citep{vmax}, taking into account that the
proper  motion cut  does not  play any  role at  all, and  binning the
luminosity function in smaller  luminosity bins (five bins per decade)
they showed that the expected statistical errors will be really small,
see Table~\ref{tab:agestats}.

These  errors  can be  considered  as  upper  bounds, as  they  depend
somewhat on  the binning, and more  precisely, on the  number of white
dwarfs in the last bin.  As can be seen the errors will be small.  The
typical  error estimate  obtained  using the  actually observed  white
dwarf  luminosity function is  1.5~Gyr, 5  times larger.   Hence, Gaia
will allow  a precise  determination of the  age of the  Galactic disk
which may  be compared  with that obtained  using other  methods, like
turn-off  stars and  isochrone fitting.   In this  case,  moreover, it
should  be  taken as  well  into account  that  Gaia  will allow  very
rigorous tests of the main sequence and red giant stellar evolutionary
models, so  additional information will be available  to constrain the
(pre-white dwarf) stellar models.

\begin{table}
\centering
\caption[Expected statistical errors in the determination  of the age of the Galactic disk]
        {Expected statistical errors in the determination  of the age
         of the disk, as obtained from fitting the cut-off in the disk
         white dwarf luminosity  function, in terms of the  age of the
         disk. See text for details.\label{tab:agestats}}         
\vspace{0.3cm}
\begin{tabular}{cc}
\hline
\hline
$T_{\rm disk}$~(Gyr) & $\Delta T_{\rm disk}$~(Gyr) \\
\hline
 8 & 0.15\\
 9 & 0.30\\
10 & 0.30\\
11 & 0.30\\
12 & 0.15\\
13 & 1.13\\
\hline
\hline
\end{tabular}
\end{table}

\subsubsection{Neutrinos}

Neutrinos are  the dominant form of  energy loss in  white dwarf stars
down to $\log(L/L_{\sun}) \simeq -2.0$, depending on the stellar mass.
As a consequence, the evolutionary timescales of white dwarfs at these
luminosities sensitively  depend on the  ratio of the  neutrino energy
loss to  the photon energy  loss, and, hence,  the slope of  the white
dwarf luminosity function directly reflects the importance of neutrino
emission.    Although  the  unified   electroweak  theory   of  lepton
interactions has been well tested  in the high-energy regime the white
dwarf luminosity function would be helpful in producing an interesting
low-energy test of the  theory. \cite{torr05} showed that the drop-off
in the white  dwarf luminosity function is not  affected by neutrinos.
However,  the  slope of  the  disk  white  luminosity function,  which
reflects the cooling rate, is sensitive to the treatment of neutrinos.
More interestingly, Gaia will be able to measure the cooling rate and,
thus, to probe the electroweak theory at low energies.

\subsubsection{Discriminating among different cooling models}

After  examining the  physical mechanisms  that operate  at moderately
high luminosities, say $\log(L/L_{\sun})\ge -2.0$, we focus now on one
of the crucial issues in the  theory of white dwarf cooling, namely on
crystallization  and  phase  separation  \citep{nature1}  at low  core
temperatures ($T_{\rm c}\sim 10^6$~K).  As discussed in \cite{Isern97}
and  demonstrated recently  in \cite{nature2}  the inclusion  of phase
separation  upon crystallisation adds  an extra  delay to  the cooling
(and, thus, considerably modifies  the characteristic cooling times at
low luminosities), which depends  on the initial chemical profile (the
ratio of  carbon to oxygen), on  the adopted phase diagram  and on the
transparency  of  the  insulating  envelope.  All  this  modifies  the
position  of  the cut-off  of  the  white  dwarf luminosity  function.
Consequently,  if a  direct measure  of the  disk age  with reasonable
precision  is obtained  by  an independent  method,  say via  turn-off
stars, Gaia will directly probe  the physics of crystallisation. It is
worth noting as well that not  only the exact location of the drop-off
of the disk white dwarf luminosity function is affected by the details
of the cooling  sequences but also, the position and  the shape of the
maximum  of  the  white   dwarf  luminosity  function,  thus  allowing
additional tests.

\subsection{Star formation rate}

The white dwarf luminosity is sensitive to the star formation history.
However, recovering the exact dependence of the star formation history
is difficult  since an inverse problem  must be solved.   In fact, the
origin of the problem is the  long lifetimes of low mass main sequence
stars.  This  implies that the  past star formation activity  is still
influencing  the  present  white  dwarf birthrate.  To  overcome  this
problem   there  exist   two   alternatives.   The   first  and   most
straightforward method requires an ``a priori'' knowledge of the shape
of  the  star formation  history  and  consists  of adopting  a  trial
function, depending  on several parameters, and searching  for the values
of these parameters that best  fit the observed luminosity function by
minimising the differences between  the observational and the computed
luminosity function.  The second possibility consists of computing the
luminosity function  of massive white dwarfs  \citep{DP94}, which have
negligible  main  sequence  lifetimes,  thus making the
solution of the inverse problem much easier.

The simulations of \cite{torr05}  indicate that a sizeable fraction of
massive  white  dwarfs(M$>$  $0.8\,
M_{\sun}$)  will be  observed by Gaia. This  fraction  varies from  7\% for  a
constant star formation  history, to 4\% for a  exponential one and to
3\% for  a episodic star  formation history --- see  \cite{torr05} for
details. These fractions are large enough to obtain the history of the
star  formation activity in  the solar  neighborhood using  the method
explained in \cite{DP94}. Although these fractions may seem small when
taken at face value, the  absolute numbers of massive white dwarfs are
impressive, since for the case  of a constant star formation rate, 700
massive white are expected to be found, whereas for the other two star
formation histories  500 and  300 massive white  dwarfs will  be found
respectively, thus allowing a determination of the luminosity function
of massive white dwarfs.

\subsection{Solar neighbourhood}

White dwarfs are relatively numerous in the Solar neighbourhood,  however,
because of their intrinsic faintness, a significant fraction that
reside within even a fairly small volume near the Sun have escaped
detection to date.  In fact, \cite{2008AJ....135.1225H} show that we
are $\sim$20\% incomplete within 20 pc and \cite{2009AJ....137.4547S}
show that we are $\sim$60\% incomplete within 25 pc.  While the
compilation methodologies differ slightly between the two works, the
results are consistent with one another as the volume is nearly
doubled when one extends from 20 pc to 25 pc.  In general, the missing
members comprise the coolest and least luminous white dwarfs.  As the
brightest representatives of this class of stars, it is imperative
that we have a complete local sample that we can then use to compile
robust statistics (e.g., multiplicity, luminosity function, mass
function, fraction with metal pollution).  Also, the coolest members
of this sample will provide empirical constraints for theory as
current atmospheric models do not provide an accurate characterisation
in this regime.

Significant efforts exist that attempt to close this incompleteness
gap.  The favored method of identifying white dwarfs is that of
reduced proper motion \citep[RPM; e.g.,][]{1972ApJ...177..245J}.
Indeed, dozens of nearby white dwarfs have been discovered with this
method \citep[e.g.,][]{2007AJ....134..252S, 2008AJ....136..899S,
2010AIPC.1273..180S}.  The inherent bias with this method, and a
significant one at that, is that it requires the candidates to have
significant proper motions in order to be identified as potential
nearby stars.  
%Given that large proper motion
%surveys typically do not extend below
%0.15$\arcsec$yr$^{-1}$ \citep{2005AJ....129.1483L,1979nltt.book.....L}, 
%the lower proper motion regime is largely unsampled.  
With the addition of synoptic photometric surveys such as
the Sloan Digital Sky Survey (SDSS), accurately calibrated photometric
magnitudes permit reliable white dwarf candidate discrimination solely
based on colour-colour selections.  However, the colours of cooler white
dwarfs ($T_{\rm eff}$ $\le$ 7,000 K) overlap with the more numerous F,
G, and K main-sequence stars and are thus photometrically
indistinguishable \citep{2006AJ....131..571H}.  It is precisely within this
temperature regime that the white dwarf luminosity function peaks (see
elsewhere in this Section) giving rise to the possibility of significant
incompleteness in a volume--limited sample in the solar neighbourhood.  In the case of the SDSS,
both astrometric and photometric data have been combined to
successfully identify white dwarfs candidates with cooler
temperatures \citep{2006AJ....131..582K, 2010AIPC.1273..180S}.  In
fact, \cite{2010AIPC.1273..180S} identified two white dwarf candidates
that were spectroscopically confirmed and likely reside within 15 pc
of the Sun (a volume known to contain only $\sim$50 white dwarfs with
accruate trigonometric parallaxes).  The SDSS covers only a fraction
of the entire sky and other synoptic surveys (e.g., Pan-STARRS, LSST)
will produce data products that permit identical analyses over larger
regions of the sky.  Yet, the intrinsic bias toward large proper
motions remains.

\begin{figure}[!t]
  \centering 
  \includegraphics[angle=90,width=0.9\textwidth]{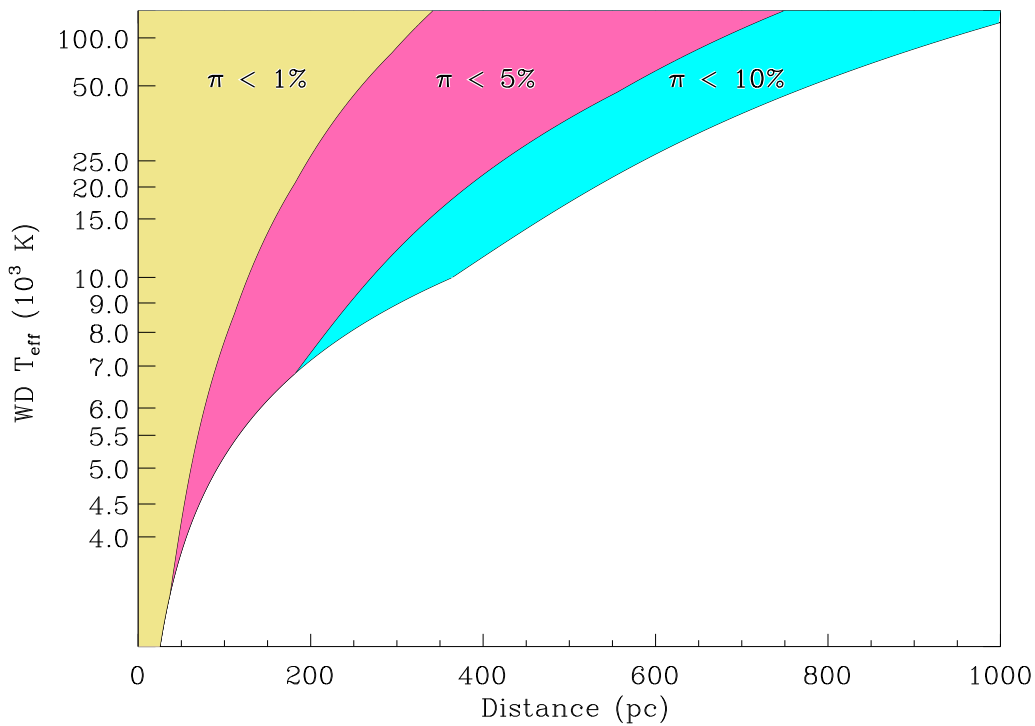} 
  \caption[Volume/completeness WD detection limits for Gaia]
  {Plot of the volume limits that Gaia can
  expect to be complete as a function of white dwarf $T_{\rm eff}$ and
  expected astrometric accuracy.  End-of-mission trigonometric
  parallax accuracies are labeled.  Expected astrometric accuracies
  for Gaia were adopted from \cite{2008IAUS..248..217L}.}  
  \label{gaiasens}
\end{figure}

To remove the proper motion bias, an all-sky survey that provides
accurate astrometry (in particular, trigonometric parallaxes) of all
stars down to a limiting magnitude is required.  Gaia will provide
exactly that and thus permit the compilation of the most complete
volume-limited white dwarf sample.  As can be seen in
Figure \ref{gaiasens}, Gaia will complete the 25 pc sample with
better than 1\% astrometric accuracies, including even the coolest
white dwarfs.  We can expect Gaia to uncover at least 150 new white
dwarfs in this volume alone.  Should Gaia uncover new white dwarfs
closer than 10 pc, our currently accepted value for the local density
changes thus increasing the number expected farther out.  While not
wholly complete, Gaia will significantly impact more distant samples
by identifying rare and exotic white dwarfs too few to be found in
smaller volume samples \citep[e.g.,][]{2011AIPC.1331..211G}.

One crucial element that Gaia will not provide is spectroscopy.
With astrometry from Gaia, white dwarf stellar types will be without
question based on their location in the H-R Diagram.  However,
accurate spectral classification (e.g., DA, DB, DC) for all white
dwarfs and accurate characterisation for exceptional white dwarfs will
require follow-up spectroscopy.  With these additional data, the
scientific return of Gaia astrometry is greatly enhanced.  For
example, metal-polluted white dwarfs are thought to have been enriched
by planetary debris disks in the very recent past or perhaps during
the current epoch \citep{2010ApJ...714.1386F}.  The implications for
planetary system remnants and possible scenarios for their evolution
as the host star evolves are great.  Yet, without the follow-up
spectroscopy, these systems appear astrometrically indistinguishable.
Additional follow-up data products include high-resolution
spectroscopy for radial velocity determinations as well as
high-resolution imaging.  Both of these data products will probe into
the question of white dwarf multiplicity that will not be available
from Gaia data alone, in most cases.

In summary, the ``Age of Astrometry'' that we are currently
experiencing is quite exciting.  Gaia will be the next profound
milestone in this regard, the full implications of which can only be
imagined.  With respect to the local sample of white dwarfs, the
quality of the Gaia data will undoubtedly illustrate limitations in
our understanding of white dwarf theory.  These empirical constraints
will, in turn, feed back into revised theories thus perpetuating a
progression of our understanding of the universe.

\subsection{Testing physical theories and new physics}

Several  non-standard   theories  predict  the   existence  of  exotic
particles.  Since  very often there are not  laboratory experiments in
the relevant energy range able  to obtain empirical evidences of their
existence  or properties,  it  is  necessary to  use  stars to  obtain
information about them  \citep{raff96}.  The general procedure adopted
in this case consists in comparing the observed properties of selected
stars, or of a family of stars, with well-measured properties with the
predictions of the theoretical stellar models obtained under different
assumptions about the underlying microphysics.  In particular, the hot
and dense  interior of stars is  a powerful source  of low-mass weakly
interacting  particles  that  freely  escape to  space.   Thus,  these
hypothetical particles  constitute a sink of energy  that modifies the
lifetimes of stars at the different evolutionary stages, thus allowing
a  comparison with the  observed lifetimes.   This technique  has been
applied  to many  of the  frontier problems  in Physics  but  the poor
quality of  the observational data  sets, among other  limitations ---
mainly  due to  our poor  knowledge of  the Galactic  populations, see
below  ---   have  only  allowed us  to  obtain  upper  bounds   to  the
characteristics of these particles,  or to obtain loose constraints to
other  hypothetical physical processes  of interest.  Typical examples
are  the bounds  to  the mass  of  the axions  \citep{raff86}, to  the
secular variation of the gravitational constant \citep{garc95}, to the
magnetic momentum  of the neutrino  \citep{blin94}, to the  density of
monopoles  \citep{free84},  to  the  size  of  large  extra-dimensions
\citep{male01} or  to the  formation of black  holes from  high energy
collisions \citep{gidd08}.

White  dwarfs  (and the  white  dwarf  populations)  can be  excellent
laboratories for testing new physics since: i) Their evolution is just
a simple cooling process, ii) The basic physical ingredients necessary
to predict their  evolution are well identified, and  iii) There is an
impressively solid observational database  to which the predictions of
the  different theories  can be  compared, that  Gaia  will undoubtely
enlarge.  In essence,  the important fact to be  used for such purpose
is simple.  Since the core  of white dwarfs is  completely degenerate,
these  stars cannot obtain  energy from  nuclear reactions,  and their
evolution is  just a gravothermal  process of contraction  and cooling
that can be roughly described as:

\begin{equation}
L_{\rm ph} + L_{\nu}  =- \frac{d(E+\Omega)}{dt}
\end{equation}

\noindent  where $E$  is the  total internal  energy, $\Omega$  is the
total gravitational energy, and $L_{\rm ph}$, $L_\nu$ and $L_{\rm es}$
are   the   photon,  neutrino   and   additional  sink   luminosities,
respectively. Therefore, the inclusion of an additional sink or source
of  energy   would  modify   the  characteristic  cooling   time  and,
consequently, the individual (for instance, the secular rate of change
of the period of pulsation of variable white dwarfs) or the collective
(namely,  the white  dwarf  luminosity function)  properties of  white
dwarfs, which are sensitive to this time scale, would be modified.

The white dwarf luminosity function  is defined as the number of white
dwarfs of a given luminosity  per unit of magnitude interval, and from
a  theoretical  point  of  view  can  be  easily  computed  using  the
expression:

\begin{equation}
n(l)\propto\int^{M_{\rm s}}_{M_{\rm i}}\,\Phi(M)\,\Psi(t)
\tau_{\rm cool}(l,M) \;dM
\label{lf}
\end{equation}

\noindent where $t = T-t_{\rm cool}(l,M)-t_{\rm PS}(M)$ is the time at
which  the progenitor  star  was born,  $l$  is the  logarithm of  the
luminosity in  solar units, $M$  is the mass  of the parent  star (for
convenience all  white dwarfs  are labeled with  the mass of  the main
sequence  progenitor),  $t_{\rm cool}$  is  the  cooling  age for  the
corresponding  luminosity, $\tau_{\rm  cool}=dt/dM_{\rm  bol}$ is  the
characteristic cooling time at his luminosity, $M_{\rm s}$ and $M_{\rm
i}$ are the maximum and the  minimum masses of the main sequence stars
able to produce  a white dwarf of luminosity $l$,  $t_{\rm PS}$ is the
lifetime of the  progenitor of the white dwarf, and $T$  is the age of
the  population under  study.  The  remaining quantities,  the initial
mass function, $\Phi(M)$, and  the star formation rate, $\Psi(t)$, are
not known  a priori and depend  on the astronomical  properties of the
stellar  population under  study.  Since  the total  density  of white
dwarfs  is not  yet well  known, the  computed luminosity  function is
customarily  normalised to  the bin  with the  smallest error  bars in
order  to compare  theory with  observations.   Equation~(\ref{lf}) clearly
shows  that in  order to  use the  luminosity function  as  a physical
laboratory it is necessary not  only to have good enough observational
data but also to know the properties of the population under study. In
particular,  for the  case of  the luminosity  function of  disk white
dwarfs the  Galactic star formation  rate, the initial  mass function,
and  the age of  the Galactic  disk must  be provided.   Actually, the
white dwarf luminosity function can be  used to obtain, or at least to
constrain,  them.  This  represents  a fundamental  limitation of  the
method,  since any anomalous  behavior of  the white  dwarf luminosity
function can be attributed to a peculiar Galactic property, and not to
a new physical phenomenon.

Nevertheless,  this situation  has recently  changed. Firstly,  it has
been  recently realised  \citep{iser08} that,  since for  large enough
luminosities the characteristic cooling time is not strongly dependent
on the mass of the white  dwarf, it is possible to cast Equation~(\ref{lf})
in the form:

\begin{equation}
n(l) \propto \langle \tau_{\rm cool}\rangle \int{\, \Phi(M) \Psi(\tau)
dM}
\end{equation}

\noindent  Hence,  if only  bright  white  dwarfs  are considered  ---
namely, those with $t_{\rm cool}  \ll T$ --- the boundary condition of
Equation~(\ref{lf})  can be satisfied  for a  wide range  of masses  of the
progenitor stars. That  is, by stars of different  masses born at very
different times.  Because of the strong  dependence of the  age of the
progenitor  on  the   mass,  the  lower  limit  of   the  integral  in
Equation~(\ref{lf})  is  almost  independent  of  the  luminosity  and  the
different  values that  takes  the integral  for  different values  of
$\Psi$ can be incorporated into  the normalization constant, in such a
way that the shape of the  bright part of the luminosity function only
depends on the characteristic cooling time.

\begin{figure}[t]
 \centering
 \includegraphics[width=0.49\textwidth]{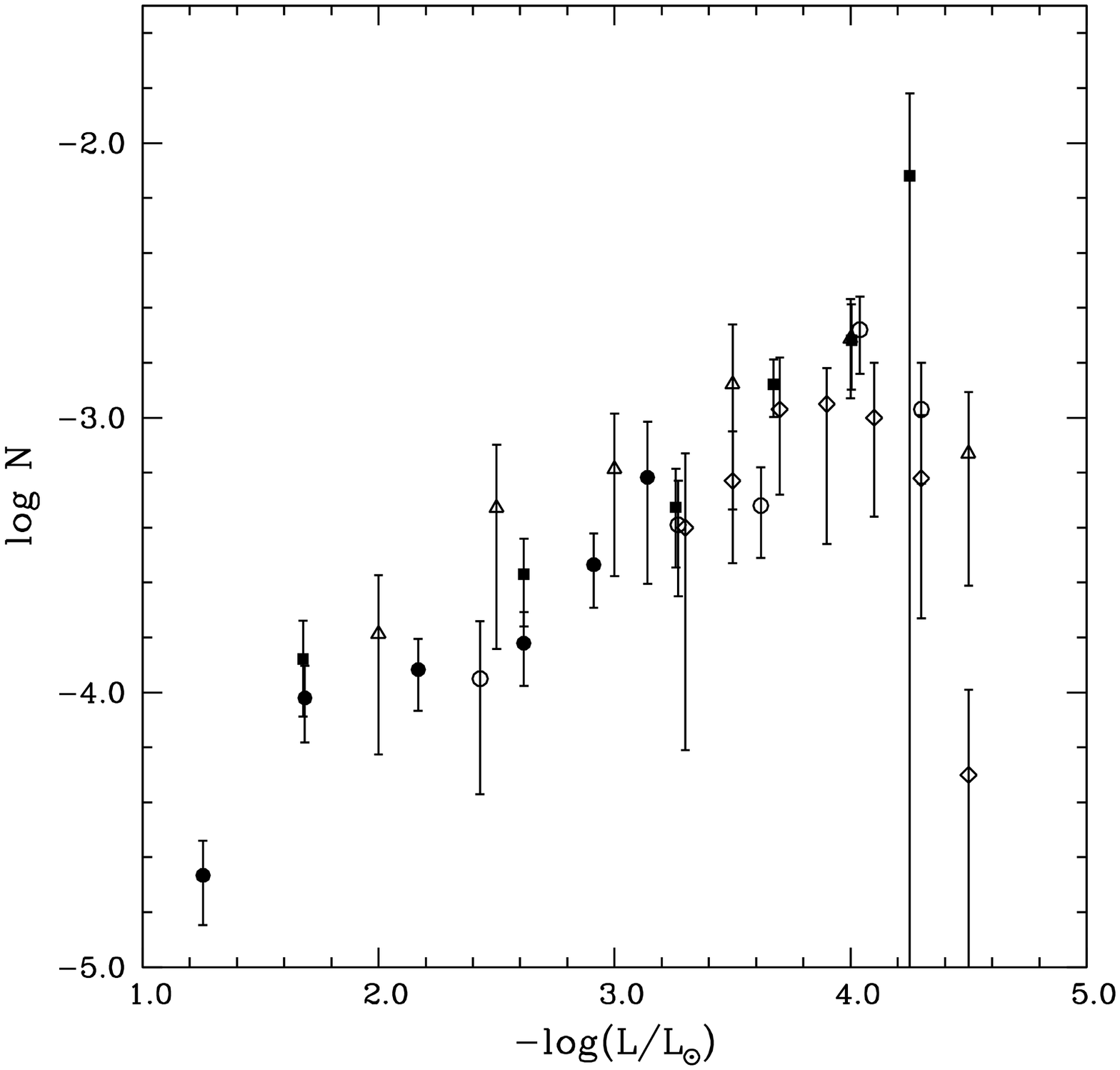}
 \includegraphics[width=0.49\textwidth]{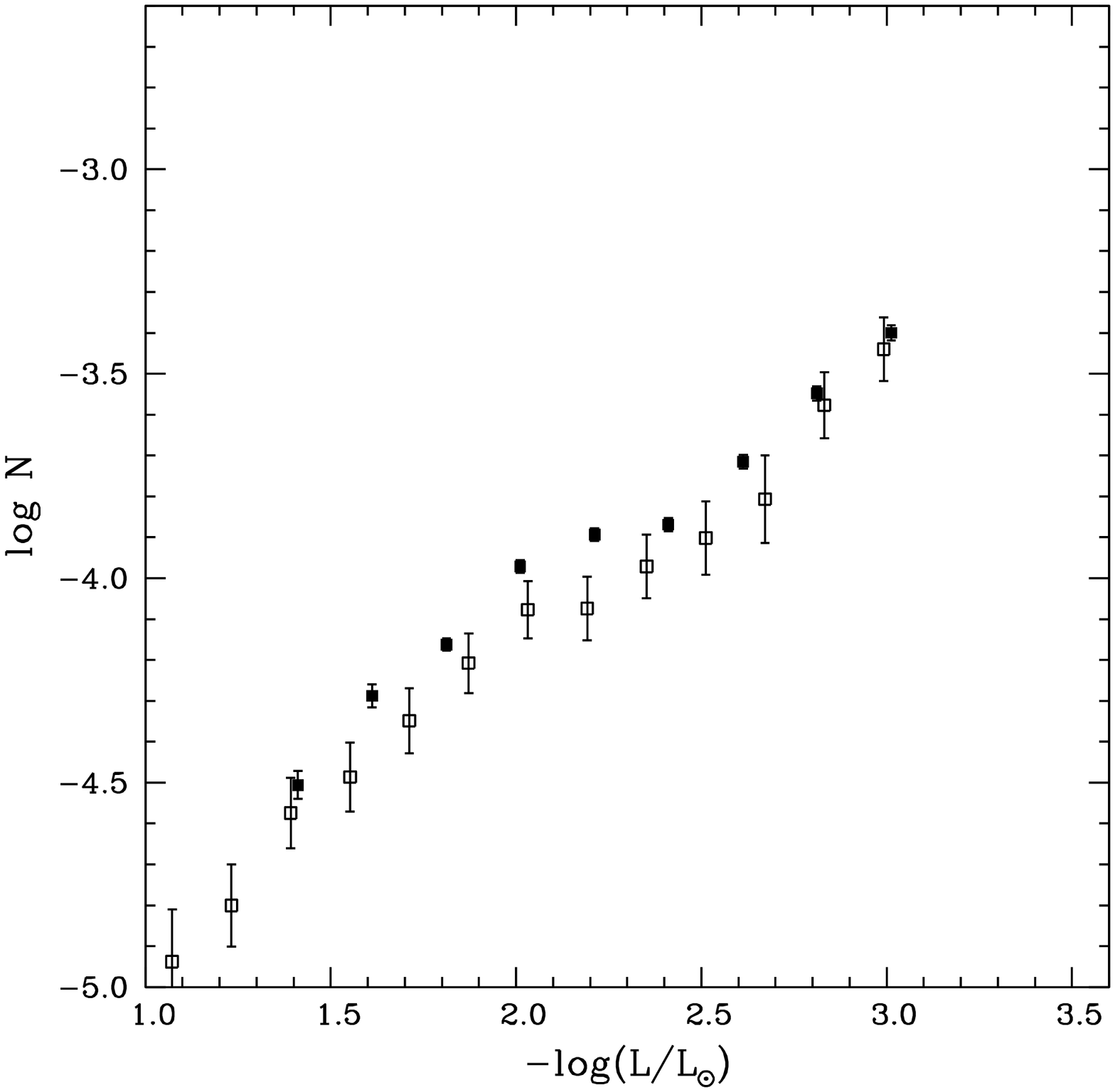}
 \caption[Luminosity Functions obtained before and during the era of large surveys]
  {Left panel: luminosity functions  obtained before the era of
   large   surveys.   The   different   symbols  represent   different
   determinations: \cite{1988ApJ...332..891L},  solid circles; \cite{evan92}, solid
   squares;   \cite{oswa96},   triangles;   \cite*{legg98},   diamonds;
   \cite*{knox99},  open circles.   Right  panel: luminosity  functions
   derived from the SDSS.  One  of them, displayed using filled square
   symbols,  was  obtained  using  both  DA and  non-DA  white  dwarfs
   identified    from    their    photometry   and    proper    motion
   \citep{2006AJ....131..571H}.  The other  one, shown  in this  figure  using open
   square symbols, was derived using only spectroscopically identified
   DA white dwarfs \citep{dege08}.}
\label{BCN.fig1}
\end{figure}

Secondly, the quality of the observational data necessary to arrive to
a definite  conclusion has also been substantially  improved, owing to
the enhanced rate of discovery of new white dwarfs of the extant large
surveys,  like  the SDSS. This  can  be
theoretically assessed in the following way.  Since the number density
of white  dwarfs at each magnitude  bin depends on the  time that each
star  takes to  cross  it, when  a  non-standard source  of energy  is
included this number changes as:

\begin{equation}
N = N_0 \frac{L_0}{L_0+L_1}
\label{bd}
\end{equation} 

\noindent  where $L_0$  is the  luminosity of  the star  obtained with
standard  physics and $L_1$  is the  contribution of  the non-standard
terms.   Consequently, the  allowed  values of  $L_1$ are  essentially
constrained       by       the       observational       uncertainties
\citep{iser08,iser08a}. The left panel of Figure~\ref{BCN.fig1} displays
the situation  just before  the beginning of  the large  surveys.  The
error bars were large and the dispersion of the different measurements
was large as well.  The right pannel of this figure shows the dramatic
change introduced  by the wealth  of observational data from  the SDSS
catalogues. The accuracy and precision  of this data has been recently
proved by  \cite{2011arXiv1102.3193R} using the data obtained  from the SuperCOSMOS
Sky Survey.

Gaia will not only improve by  two orders of magnitude the size of the
current samples, but  it will also allow to  determine the position of
the  age   cutoff,  which  is   basic  for  some   applications,  like
constraining  the  secular   variation  of  the  gravitation  constant
$G$. Additionally,  it will also  allow to disentangle in  an absolute
way  the  influence  of  the  star formation  rate  by  obtaining  the
luminosity function of massive white dwarfs \citep{DP94,torr05}.

\subsubsection{The case of axions}

One solution to the strong CP problem of quantum chromodynamics is the
Peccei-Quinn  symmetry   \citep{pecc77a,pecc77b}.   This  symmetry  is
spontaneously  broken  at an  energy  scale  that  gives rise  to  the
formation   of   a    light   pseudo-scalar   particle   named   axion
\citep{wein78,wilc78}.  This  scale of energies is not  defined by the
theory but  it has to  be well above  the electroweak scale  to ensure
that the coupling between axions  and matter is weak enough to account
for the lack  of positive detection up to now. The  mass of axions and
the  energy  scale  are  related  by: $m_{\rm  a}  =  0.6(10^7\,  {\rm
GeV}/f_{\rm  a})$~eV.  Astrophysical  and cosmological  arguments have
been used to  constrain this mass to the range  $10^{-4} {\rm eV} \leq
m_{\rm  a} \leq  10^{-4}~{\rm eV}$.  For this  mass range,  axions can
escape from  stars and  act as  a sink of  energy. Therefore,  if they
exist, they can noticeably modify the cooling of white dwarf stars.

Axions can couple  to photons, electrons and nucleons  with a strength
that  depends  on  the  specific implementation  of  the  Peccei-Quinn
mechanism.   The   two  most  common  implementations   are  the  KSVZ
\citep{kim79} and the DFSZ  models \citep*{dine81}.  In the first case,
axions couple  to hadrons and photons,  while in the  second they also
couple to charged  leptons. For the temperatures and  densities of the
white  dwarf  interiors  under  consideration, only  DFSZ  axions  are
relevant,  and in  this  case they  can  be emitted  by Compton,  pair
annihilation and bremsstrahlung processes, but only the last mechanism
turns out to be important.

\begin{figure}[t]
 \centering
 \includegraphics[width=0.8\textwidth]{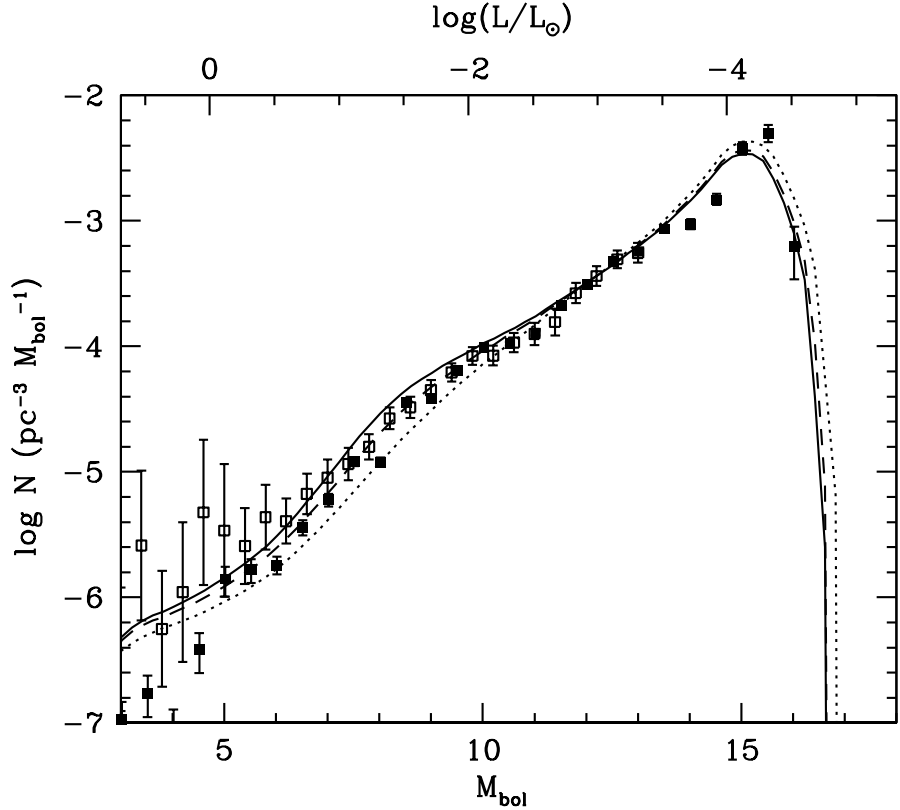}
 \caption[White dwarf luminosity functions for different axion masses]
  {White dwarf luminosity functions for different axion masses:
   $m_{\rm a}\cos^2\beta$  = 0  (solid line), 5  (dashed line)  and 10
   (dotted line) meV.}
\label{BCN:fig2}
\end{figure}

Figure \ref{BCN:fig2}  displays the luminosity  function for different
axion  masses,  a constant  star  formation rate  and  an  age of  the
Galactic  disk of  11 Gyr.   All  the luminosity  functions have  been
normalized to  the luminosity bin at  $\log (L/L_{\odot})\simeq-3$ or,
equivalently, $M_{\rm bol} \simeq 12.2$.  The best fit model -- namely
that  which  minimises  the  $\chi^2$  test  in  the  region  $-1>\log
(L/L_{\odot})>-3$ (that is, $ 7.2 < M_{\rm bol} < 12.2$), which is the
region where  both the observational  data and the  theoretical models
are reliable  -- is obtained  for $m_{\rm a} \cos^2  \beta\approx 5.5$
meV and solutions  with $m_{\rm a} \cos^2 \beta >  10$ meV are clearly
excluded \citep{iser08a,iser09}.   Taken at face  value, these results
not only  provide a strong constraint  on the allowed  mass of axions,
but also a first evidence of their existence and a rough estimation of
their mass.   This is of  course a strong  statement but it is  also a
typical example of the problems to  be solved or posed with the future
white dwarf luminosity functions that Gaia will provide.

\subsection{IMF}
 
The initial mass function (IMF) is the distribution of stellar masses at birth. It is a fundamental property that 
quantifies the efficiency of the conversion of gas into stars in galaxies and thus, it is needed to understand many 
fundamental astrophysical problems: star formation in galaxies, chemical evolution and nucleosynthesis, supernova 
rates in galaxies, models of galaxy formation and evolution...

The initial mass function is usually expressed as $\frac{dN}{d(M)} \propto M^{-\alpha}$, or in its logarithmic 
form, $\frac{dN}{d(\log(M))} \propto M^{-\Gamma}$, where M is the stellar mass. Up to now, most authors have been using what is 
called the standard initial mass function, which was determined by Salpeter more than 50 years ago \citep{1955ApJ...121..161S}, 
where $\Gamma=-1.35$. However, \citet{1986IAUS..116..451S} noticed that a single power law cannot reproduce the shape of the IMF 
over the full stellar mass range, since there was a strong rollover at low masses and a steepening of the function 
relative to Salpeters at masses higher than $1{\rm M}_{\odot}$. Subsequent studies have confirmed the decrease in 
slope at low masses obtaining a multisegment power law which is often used in the literature \citep{1993MNRAS.262..545K}.

There are several fields within Astronomy and Astrophysics that have the aim of understanding the formation of the 
Universe and our Galaxy. One way is to observe galaxies similar to ours but located at large distances, although 
the stellar populations are not fully resolved and to measure star formation rates requires the adoption of an 
initial mass function, with some authors favouring top-heavy prescriptions \citep[e.g.][]{2005MNRAS.356.1191B}.  A direct test is 
possible locally in our Milky Way by studying stars of the halo and thick disk populations, which were formed at 
the same age of the Universe as the starbursts observed in the distant galaxies. The evolution of white dwarfs is 
driven by a simple cooling process, which has a duration of the order of the age of the Galaxy \citep{2000ApJ...544.1036S}. Thus, 
they have imprinted a memory of the various episodes that the Galaxy has been subject to over its history, 
constituting useful objects to probe its structure and evolution (\citealt{2001ApSSS.277..273I};\citealt*{2005ApJS..156...47L}).

Although all population II stars more massive than the Sun evolved away from the main sequence long ago, the early 
IMF can be reconstructed from the luminosity function of the relic white dwarf population. The thin disc luminosity 
function is a complicated convolution of the star formation history of the Galaxy, the IMF, the initial-final-mass 
relationship, secular evolution of the thin disk scale height and other effects. A deconvolution is thus extremely 
difficult. However, the interpretation of the thick disc/halo luminosity function is much more straightforward. The 
formation history of the halo and thick disc was probably complex, but observations of field stars and globular 
clusters indicate that most stars are very old. So we can assume that both halo and thick disc were formed over a 
short period of time with no further star formation afterwards. Higher mass stars evolved to the white dwarf stage 
after only a short amount of time and can now be found piled up at the cool end of the white dwarf luminosity 
functions. Thus, it is essential to detect and study cool white dwarfs in order to be able to study the oldest 
population of white dwarfs, and in this way obtain information about the initial start formation bursts.

\citet{2001Sci...292..698O} reported a very large population of white dwarfs belonging to the galactic halo -- in line with a top 
heavy IMF. This result was disputed by \citet*{2001ApJ...559..942R} and \citet{2006AandA...447..173P}, but all these investigations are based on 
very local samples and large uncertainties remain, because of small number statistics. A particular problem is the 
lack of confirmed very cool population II white dwarfs, which are most interesting, because they evolved from the 
most massive progenitor stars. Even the population I white dwarf luminosity function constructed by \citet{2006AJ....131..571H}, 
comprising of 6,000 WDs selected from the SDSS contains only 35 cool WDs with absolute 
magnitudes $>15$\,mag. In a recent work, \citet{2010ApJS..190...77K} identified few cool white dwarfs using SDSS data, although only a couple have proper 
motions compatible with the halo population.

\citet{2009JPhCS.172a2004N} constructed a model of the Galactic WD population, based on the model of Galactic structure by 
\citet{2003AandA...409..523R}. These simulations consider the WD cooling sequences from \citet{1995AandA...299..755B}, the stellar tracks of 
\citet{2000AandAS..141..371G}  and the initial-final mass relationship of 
\citet{2001ASPC..226...13W}. The Salpeter IMF is assumed, although the simulations can be run with other IMFs \citep{2005MNRAS.356.1191B,1983ApJ...272...54K}. 
The population identification is then based on the results of the kinematic study of \citet{2006AandA...447..173P}, 
calibrated with the local sample \citep{2008AJ....135.1225H} and checked against the proper motion selected sample of cool WDs by 
\citet{2001Sci...292..698O}. According to these simulations 2552, 1655 and 444 white dwarfs with $T_{\rm eff}<4000$K belonging to 
the thin-disc, thick disc and halo, respectively, will be detected by Gaia. Gaia will also provide accurate trigonometric parallaxes and proper 
motions that will guarantee a correct population identification and will allow a considerable increase of the census of the halo white dwarf 
population.

% We 
%have used the synthetic white dwarf photometry from \cite{2006AJ....132.1221H} and the transformation equations kindly supplied by 
%the Gaia team to convert SDSS magnitudes into Gaia magnitudes. 
Gaia performs low-resolution spectroscopy in four different bands, which can be rebinned into photometric bands.
Comparing SDSS magnitudes in the Gaia filter set ($G_{BP}$ and $G_{RP}$) to these synthetic Gaia magnitudes, a rough estimate of $T_{\rm eff}$ can be obtained for white dwarfs. 
The photometric spectral energy distribution from Gaia also is sensitive to different compositions at low temperatures, meaning it is possible to discern 
between H or He-rich atmospheres. However, IR photometry is also required for a complete characterisation (accurate $T_{\rm eff}$ 
and $\log g$)of the white dwarfs detected. 
%(accurate $T_{\rm eff}$ 
%and $\log g$). This information could be obtained from some IR surveys like the UKIDSS Large Area Survey (LAS), 4,000 
%sq.~deg.~of coverage and limiting magnitude $J=$20; VISTA VIKING, 1,500 sq. deg. and $J=$22.1; and VHS (Vista 
%Hemisphere survey), 20,000 sq.~deg.~and $J=$21.2.

Even though Gaia is not very deep, $G=$20, it will cover the whole sky which guarantees the detection of cool, nearby white 
dwarfs. For instance a 3,000K white dwarf will 
be detected by Gaia only if its closer than 44 or 31 pc, for $\log g$=8.0 or 8.5, respectively. Once we have a large 
sample of cool white dwarfs a more comprehensive halo white dwarf luminosity function  will be 
obtained. This will also enable us to infer which of the initial mass functions in the literature is 
the most realistic one.

%\subsection{Galactic populations of white dwarfs}

\newpage

\putbib[LF_refs]

%% file: section_mag_fields.tex
\section{Magnetic fields (origin and evolution)}

Magnetic fields have been measured in approximately 200 white dwarfs with field
strengths of 10\,kG$<$B$<$1000\,MG. The Sloan Digital Sky Survey (SDSS) has been
important in discovering new magnetic white dwarfs: where the spectroscopic data
has been particularly important for identification and determination of magnetic
field strengths through spectral fitting along with other useful parameters,
such as the temperature.

There are multiple hypothesis for the origin of these objects \citep[see e.g.][]{KulebiPhd}:  
One of them, the ``fossil  field'' hypothesis suggests that magnetic fields are products of an earlier stage of stellar
evolution. In this picture, the field strengths are amplified due to the contraction of the
core, during which the magnetic flux is conserved to a major extent. From the perspective
of this hypothesis, chemically peculiar Ap and Bp stars were proposed to be the progenitors
of MWDs (Angel et al., 1981).

The ``fossil field'' hypothesis has certain problems, the most important one being the
incommensurable incidence of the magnetism in different stages of the stellar evolution.
Namely, the inferred incidence of MWDs with respect to the total white dwarf population
outnumbers the incidence of Ap/Bp stars within the A/B population \citep{KAW07}.
One other problem of the fossil field hypothesis is the relatively massive nature of the MWDs
\citep{Liebert88,VennesKawka08}. While the mean value of the masses of the MWDs
is $\approx  0.93 M_{\odot}$, the mean mass of the non-magnetic white sample is $\approx 0.56 M_{\odot}$ \citep{Liebert88}.
It has been suggested that this could be a result of the
influence of magnetism on the mass loss. This possibility was tested by \citet{WickramsingheFerrario05} via
population synthesis. Their conclusion was that current number distribution and masses of
high-field magnetic white dwarfs (HFMWDs, $B\le 106$ G) are not mainly due to an inclusion
of a modified IFMR but rather by assuming that $\approx$∼ 10\%\ of A/B stars have unobservable
small scale magnetic fields.

Magnetism is thought to be
present in more than 10\,\% of white dwarfs in the local population
(\citealt{LIE03}), with \cite{KAW07} finding the incidence of magnetism as
21\,$\pm$\,8\% within 13\,pc.

Accurate parallax and distance measurements for targets will mean estimates for
the radius of objects can be obtained -- a parameter which is difficult to
determine accurately. 
The proper motion information from Gaia could help to reveal the origins of
magnetic fields in white dwarfs. In a study by \cite{ANS99}, the statistical
properties, cooling ages and vector components of the three-dimensional space
motions U, V, W were examined for a sample of 53 magnetic white dwarfs in the
fourth edition of the Catalog of Spectroscopically Identified White Dwarfs
(\citealt{McCook99}). They concluded that the sample of magnetic white dwarfs
appeared to descend from young disk stars, due to their small motions relative
to the sun. This supports the hypothesis that magnetic fields in white dwarfs
could be remnants from magnetic fields in main-sequence stars, such as Ap/Bp
stars. However, they also suggested that the magnetic white dwarf population
could have a mixture of progenitors, as no difference in velocity dispersion was
detected between the hotter (with shorter cooling ages) and cooler magnetic
white dwarfs (with longer cooling times), which would be expected if the
magnetic white dwarf population descended solely from young disk stars. The
information acquired from the Gaia mission will provide unprecedented parallax,
distance and proper motion accuracy facilitating similar studies to be carried
out with much larger sample sizes to allow thorough statistical analysis.  

The accurate luminosities obtained from the photometry will yield some
information regarding the variability of the magnetic white dwarf population.
Periods of photometric variability may not be determined from the Gaia
observations alone, however it will indicate those showing signs of modulations
to be followed-up later from ground-based facilities. It is currently thought
that $\sim$40\,\% of magnetic white dwarfs show signs of photometric variability
(\citealt{BRI07}), revealing information regarding the rotation properties of
the object. Recently, a unique variable magnetic DA white dwarf was discovered
in the Kepler field by Holberg \& Howell (2011, in prep.). The object was found
through its non-sinusoidal modulations of 4.92\,\% peak-to-peak with a period of
0.2557\,days (6.138\,hrs). With the absence of evidence to suggest that there
may be a companion or pulsations, it illustrates that the photometric
variability exhibited by some magnetic white dwarfs can be used as a method for
identifying the objects.  

Ground-based follow-up spectroscopy of these objects will be crucial to getting
the most out of the Gaia observations. In combination with spectral fitting and
evolutionary models, information regarding the object's temperature, magnetic
field strength, composition and cooling age can be determined. Estimates will
also be possible for the mass and radius of the object, although these
parameters will depend on the evolutionary models. 

Sophisticated models for the analyses of such follow-up observations could be used to
determine magnetic field strengths and estimation for the magnetic field geometry as was
demonstrated e.g. by \citet{Kulebietal09} for the SDSS data.

One of the obstacles in analysing  magnetic white dwarfs is that surface gravities cannot
reliably be determined from the spectral lines. This is due to the lack of atomic data in the presence of both a magnetic and an electric field 
for arbitrary strength and arbitrary angles between two fields. Therefore, only a crude approximation  \citep{Jordan92} is used in our model and
systematic uncertainties are unavoidable, particularly  particularly in the low-field regime ($\le 5$\,MG) where the Stark effect dominates. 
Therefore, mass can only be estimated if the determination of the effective temperature (which is more reliably possible) is supplemented by
parallax determinations.

Work conducted by \cite{KUL10} demonstrates what can be achieved with accurate
parallax measurements, follow-up spectroscopy and theoretical modelling. They
observed the highly magnetic, hot and ultra massive white dwarf RE\,J0317-853
with HST to determine the parallax of the object and its DA white dwarf
companion. The mass, radius and cooling ages were then calculated for a range of
temperatures and surface gravities using carbon-oxygen core white dwarf cooling
models with thick hydrogen layers
(\citealt{WOOD95,HOL06})\footnote{
http://www.astro.umontreal.ca/~bergeron/CoolingModels} and oxygen-neon white
dwarf cooling models with thin hydrogen layers
(\citealt{ALT05,ALT07})\footnote{
http://www.fcaglp.unlp.edu.ar/evolgroup/tracks.html}. 
In the case of RE\,J0317-853 this information together with an analysis of its companion 
LB\,9802 allowed to explore different evolutionary scenarios for RE\,J0317-853  (single-star, binary origin).   

Since Gaia is expected to find $\approx$400,000 white dwarfs \citep{Torresetal05,jordan2007} that a few percent 
 will be identified as being magnetic by follow-up observations. By analysing this sample with model spectra
 we can obtain a much clearer picture of this population. We can test whether the masses of
 magnetic white dwarfs are indeed larger than for non-magnetic white dwarfs \citep{Liebert88}, whether  there is a dependence
of the magnetic field strength with age (using the effective temperature and mass). With the help of population synthesis models 
we will be able to  further constrain the question of the origin of magnetic white dwarfs. 

Of particular importance is the studies of magnetic whites in wide binaries \citep{KUL10}, common-proper-motion systems \citep{Girvenetal10}, and open clusters
because these systems provide additional age constraints and allow the study of the evolution of these objects in great detail.

\putbib[mag_refs]

%% file: section_pulsations.tex
\section{Pulsating white dwarfs}

White dwarf (WD) asteroseismology is a powerful tool to study WD interiors.
It allows the measurements of various stellar parameters, such as stellar mass, the mass 
of the H/He external layers and the stellar rotation rate. It can also 
provide precious information on the core chemical composition, neutrino 
(and axion) cooling, and even crystallization for massive white dwarfs.
Wide review papers on WD asteroseismology are from \cite{}, \cite{}, and 
\cite{}.
Along their cooling sequence, there are four instability strips at various effective temperatures. Thus asteroseismic studies allow detailed study of 
white dwarfs at different effective temperatures and different evolutionary 
phases.
The ZZ Ceti (or DAV) instability strip is by far the most populated one, with about 150 DAV stars known \citep[][]{castanheira+2010}.

\subsection{Photometric finding of ZZ candidates}
ZZ Ceti stars are pulsating white dwarfs with effective temperatures close to the maximum of the strengths of the Balmer lines.
For this reason even at about the Gaia magnitude $G\approx 19$ Balmer lines can be identified if all BP low-resolution spectra 
are added up.  These objects can be regarded as ZZ Ceti candidates and checked for photometric variability. Moreover, they can be
tested for periodicity as described in Sect.\,\ref{sec:directperiod}.

\subsection{Direct Period detection with Gaia}
\label{sec:directperiod}
\citet{Varadietal2009} and  \citet{Varadietal2011} have performed studies to determine whether Gaia can directly measure the periods of ZZ Ceti stars.
The Gaia time sampling and the CCD data
acquisition scheme allow, in principle, the probeing of stellar
variability on time scales even as short as several tens
of seconds, thereby giving potential access to the study
of variable stars in a large and homogenous sample of
stars. In order to test this, 
continuous ZZ Ceti light-curves were simulated with  a nonlinear ZZ Ceti (DAV) light-curve
simulator based on a model by \citet{Schlund}.
The model takes periods, amplitudes and phases as
input to define a pulsation pattern at the base of the
convective zone and solves the problem of flux
propagation through the convective zone to derive the
relative flux variations at the photosphere.

In order to simulate photometric Gaia time series of
ZZ Ceti stars, the continuously
simulated ZZ Ceti light-curves were sampled according to the
nominal scanning law of Gaia assuming different positions of the
objects on the sky. Subsequently 
Gaussian noise was added with a level consistent with the 
precision of the Gaia satellite.

It turned out that at the Gaia magnitude $G=15$, periods could be partially recovered with a 
probability of $72\%$ if the power spectrum does not change during the mission. If the pulsation modes change
once during the mission the recovery probability drops to about 50\%. Even at $G=18$ a substantial percentage of 
periods is recovered.

\putbib[pulsating_refs]

%% file: section_binaries.tex
\section{Binaries}
White dwarfs in binary systems provide unique opportunities to study the basic properties of degenerate stars such as masses and radii.  They also offer excellent 'laboratories' to probe key astrophysical processes like the initial-final mass relation, the accretion of material onto white dwarf surfaces and the exotic outcomes of close binary evolution including, neutron stars and black holes.    A census of the local white dwarf population within 20pc \citep{holberg08, holberg09} indicate about 30\% of white dwarfs occur in binary systems.  This can be broken down into roughly  12\% occurring in white dwarf-red dwarf pairs, 8\% in Sirius-Like systems and 5\% + 5\% in double degenerate systems.  In this section the use of the term 'binary' includes non-interacting systems which may contain more than just two components.

\subsection{Finding new binaries}
It was due to the subtitle proper motion perturbations of the nearby bright stars Sirius and Procyon that Friedrich Bessel in 1844 first deduced the existence of unseen 'dark stars', massive enough to gravitationally cause 50 and 40 year periodic sinuous shifts in the proper motions of these stars.  The white dwarf companions to Sirius and Procyon were finally observed in 1862 and 1896, respectively, using the largest telescopes of that era.    However, It remained until 1913 before the truly unusual properties of these low luminosity yet massive companions were widely appreciated.  It was the faint but bluish companion of the K-star 40 Eridani  that independently attracted the attention Henry Norris Russell and Ejnar Hertzsprung and made clear the these faint companions were nothing like the familiar  'giant' or 'dwarf'  stars they were studying.   The name 'white dwarf' finally was coined in 1922 as an appropriate description of these blue subluminous stars.  In all of the above it was the fact that these white dwarfs were paired with bright nearby luminous stars in binary systems that first attracted the attention of astronomers.
Since that time tens of thousands of white dwarfs have been discovered by various methods form proper motion surveys to photometric surveys that turn up faint blue stars.  Most of these white dwarfs are not members of binary systems but new and interesting binary stars continue to turn up.   White dwarfs in binary systems reveal themselves in various ways.  Common proper motion (CPM) pairs are a frequent source of such discoveries.   Indeed many of the resolved binary systems in the \citet{mccook99} White Dwarf Catalog were found through the long term efforts of Willem Luyten and are contained in \citet{oswalt88}.  New systems continue to be found by this method. For example by searching for widely separated binary components associated with the Hipparcos sample \citet{gould04} have uncovered a number of new wide binary systems.  
For unresolved binary systems the white dwarf can be more luminous in the optical than the main sequence companion and thus the companion can easily detected photometrically or spectroscopically as a red excess object.  This is particularly true for M dwarf companions.  Large numbers of these systems are now known for example from the SDSS survey \citep{silvestri06}.  For more luminous main sequence companions, earlier than M) unresolved white dwarfs are difficult to detect in the optical.   However, space-based UV and EUVE observations have revealed many such systems, where the existence of either a shortwave length (UV, EUV or soft X-ray) excesses associated with main sequence stars.  In some cases these excesses were accidentally discovered; for example the chance discovery of a characteristic white dwarf UV continuum in an IUE spectrum.  But most prolific source of such systems has turned out to be the ROSAT and EUVE all sky surveys where EUV excesses or continuum where often revealing the presence of a hot white dwarf secondary star \citep{burleigh98}.  
One area where Gaia can contribute significantly is through the discovery of additional Sirius-Like systems, where a white dwarf is paired with a luminous star of spectral type K or earlier.   Such systems can be very to difficult to recognize  if the stars are too close to be resolved.   For example, Holberg, et al. (2012, in preparation) show that beyond 20 pc  anywhere from 80 to 90\% of Sirius-Like systems remain undetected.  
There are several ways that Gaia can detect new binary systems containing white dwarfs.  Astrometrically Gaia will ultimately be able to identify most common proper motion pairs by matching proper motions and distances.  For systems where the white dwarf remains hidden in the glare of the primary (Sirius-Like Systems) Gaia may detect asterometric accelerations in the primary.   Finally, Gaia can also detect tell-tale photometric variations due either to a reflection effect or eclipse variations. 

\subsection{Common proper motion binaries}
The photometric depth and proper motion precision possible with Gaia ensures that most common proper motion binaries containing white dwarfs will be found to distances of at least several hundred parsecs.  Gaia's precise global astrometric capabilities will identify any remaining undetected systems with separations from tens to hundreds of arc seconds based on common proper motions alone; however, the precise trigonometric parallaxes at the 50 $\mu$as level will also provide common distances as an additional  identifier of CPM systems.  Likewise, Gaia's dispersive photometric system will permit the identification of WDs and even first order determinations of WD spectral types \citep{garcia05}.   Holberg et al. (2012) show that for all currently resolved Sirius-Like systems, regardless of separation or orbital period, Gaia will be able to measure orbital motions in white dwarf-primary pairs during its five-year prime mission. 
\subsection{Unresolved Systems}
Of more specific interest for unrecognized Sirius-Like systems will be Gaia's ability to identify potential unresolved systems through recognition of non-rectilinear motions of the luminous star, in much the same way that the existence of the companions of Sirius and Procyon were first discovered nearly two centuries ago.  Considering the current set of Sirius-Like systems as a template and using  a proper motion acceleration of 8 $\mu$as yr$^{-2}$  as a threshold, Holberg et al. (2012) show that for many systems with known masses where the orbital separations are known or can be estimated from observed orbital periods Gaia will detect the reflex motions of the primary star over its five-year primary mission.

\putbib[bin_refs]

%% file: section_planets.tex
\section{Planetary systems}

Current exoplanet research is strongly oriented towards main sequence (MS) 
stars, with the main goal of finding rocky analogues of our Earth, and towards 
planetary system formation, while the late-stage evolution after the MS remains 
poorly understood. 
Indeed, the large majority of the almost $\approx$550 extrasolar planets 
detected so far orbit MS stars, and these planets were detected mostly using 
radial velocities (RVs) or transits. 
Both methods present strong limitations for compact stars with high gravities 
and small radii, such as the white dwarfs and their progenitors, including the 
extreme horizontal branch (EHB) stars.
This is why most of the ongoing RV or transit searches, including CoRoT and 
Kepler, will not add much to this topic.
For compact stars more efficient techniques are given by timing and astrometry.
Both methods, unlike RVs and transits, are sensitive to planets with orbital 
distances of the order of 1 AU or more.
This, in some sense, corresponds to the general expectation that very close 
planets can barely ``survive'' the most critical phases of stellar evolution 
like the red giant branch (RGB) expansion, the asymptotic giant branch (AGB) 
expansion, and the planetary nebula (PN) ejection.
To model these phenomena is difficult: the first attempts are very recent
\citep{villaver+2007, villaver+2009, nordhaus+2010}
and lack observational constraints.
For these reasons  increasing the poor statistics of these rare systems is 
important.

Presently, only few post-RGB planetary systems are known 
(see \citealt{silvotti+2011} for an updated list):
apart from the pulsar planets, they were all discovered in the last 3 years and 
were almost all detected using the timing method. 
Their host starts are EHB stars or cataclysmic variables.
For single white dwarfs, we know of only two planet candidates:
orbiting GD66 \citep{mullally+2008, mullally+2009} and GD356 
\citep{wickramasinghe+2010}, both of which are not yet confirmed.
At the same time we know that at least 21 white dwarfs show dusty/gaseous 
circumstellar disks, likely due to tidal disruption of asteroids
\citep[see e.g.][]{farihi2011, gansicke2011}.
% (see section \ref{} for more details).

In other terms, we do not know yet whether white dwarfs have planets or not 
and at which orbital distances, but we see debris disks which are 
likely the remnants of old planetary systems.
Moreover, potentially, white dwarfs could host not only 1st generation planets 
formed from the original protoplanetary disk, but also 2nd or even 3rd 
generation planets formed more recently from the stellar material ejected 
during the RGB, AGB or PN phase \citep[see e.g.][]{perets2011}.

In this context it is evident that Gaia can play an important role.

%\subsection{Debris disks}
%
% I (ROBERTO) PROPOSE TO REMOVE THIS SECTION AND POSSIBLY ADD SOME MORE DETAIL 
% ON THIS SUBJECT IN THE PREVIOUS SECTION 
% This is fine by me (Sarah)

%\subsection{Detection by astrometry and transits}
%
% I PROPOSE TO SPLIT THIS SECTION IN 2

\subsection{Detection by astrometry}

\begin{figure}[h!]
\begin{center}
\includegraphics[scale=0.5,angle=0]{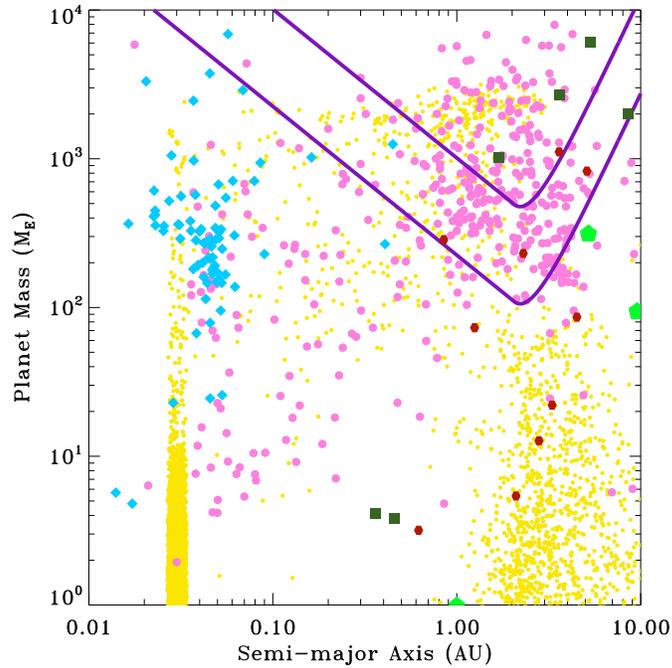}
\caption[Exoplanet discovery space for the Gaia mission based on double-blind
tests.]{Exoplanet discovery space for the Gaia mission based on double-blind
tests \citep{casertano+2008}.
Detectability curves are defined on the basis of a 3$\sigma$ criterion for 
signal detection. 
The upper and lower curves are for Gaia astrometry with 
$\sigma_A$ = 15 $\mu$as (where $\sigma_A$ is the single-measurement 
astrometric precision), assuming a 0.59-M$_{\odot}$ WD primary at 100 pc 
(V$<$15) and 50 pc (V$<$13), respectively. 
Survey duration is set to 5 yr. Pink dots indicate the inventory of 
Doppler-detected exoplanets as of May 2010. 
Transiting systems are shown as light-blue filled diamonds, while the red
hexagons and the dark-green squares are planets detected by microlensing or 
timing respectively.
When the inclination of the system is not known, in particular for the 
pink dots and the dark-green squares, we used the minimum mass. 
Solar System planets are shown as light-green pentagons. 
The small yellow dots represent a theoretical distribution of masses and 
final orbital semi-major axes from \citet{ida+2008}.}
\label{fig}
\end{center}
\end{figure}

The astrometric accuracy of Gaia is enough to detect giant planets around 
nearby white dwarfs.
Preliminary calculations have shown that Gaia should be able to detect WD 
planets around $\approx$10$^3$ white dwarfs within $\sim$100 pc from the Sun, 
with low mass limits of a few Jupiters \citep*{silvotti+2011b}.
The exoplanet discovery space of the Gaia mission is represented in 
Fig.\,\ref{fig}.
More detailed simulations will be done in the next months in order to evaluate 
the Gaia sensitivity with a real sample of white dwarfs.

\vspace{8mm}

%\subsection{Detection by transits}
% TBDone (probably by Matt) if the numbers tell us that there is a concrete 
%possibility to find something ...

\putbib[planet_refs]

%% file: section_late_stages.tex
\section{Gaia and other late stages of stellar evolution}

Gaia's \citep{1996A&AS..116..579L}
role in exploring the end states of stellar evolution is not
limited to the study of white dwarfs, but extends to many stellar
types which are their immediate progenitors or (sometimes) progeny.
Many of these are rare and distant; Gaia will provide the first
opportunity to measure proper motions, and distances for many, as well
as radial velocities and photometry of unprecedented precision
\citep{2009MmSAI..80...97C}. These
data will help to complete a picture of stellar evolution which is
severely complicated by the diverse consequences of interaction
between two stars in a binary system.

Although all stars commence evolution burning hydrogen in their cores,
with lifetimes ranging from $<10^6$ to $>10^{10}$ years, their
subsequent fate is strongly correlated with their mass.  Excepting the
comparatively rare massive stars, most stars develop a degenerate
helium core as they become red giants.  In some circumstances, such a
star can lose almost its entire hydrogen envelope while its helium
core is still small; then helium cannot ignite and a 
{\bf helium-core WD} results.
If mass-loss from the red giant is gradual, the helium
core will grow until conditions for helium ignition are achieved, some
$10^{9}$ years after leaving the Main Sequence 
(for a $1.0 {\rm M_{\odot}}$ star).
A helium flash occurs, the degeneracy of the core is lifted, core
helium-burning is established, the core expands, the hydrogen-burning
shell is cooled, the total radius drops, and a horizontal branch star
results, with a typical lifetime of $10^8$ years 
(for a $0.5 {\rm  M_{\odot}}$ helium core).

The radius and future evolution of a horizontal branch star turns out
to depend on how much hydrogen remains around the helium core.  In
extreme cases and usually in a binary system, a red giant may lose
nearly all of its hydrogen envelope at or just before the helium
flash. The resulting horizontal-branch star is virtually a helium
main-sequence star, lying well to the blue of the hydrogen main
sequence.  Such stars are identified with {\bf subdwarf B stars}.  For
these and other blue horizontal-branch stars, the absence of a
hydrogen envelope means that, as core-helium burning is followed by
shell-helium burning, possibly as a {\bf subdwarf O star}, the
available fuel store is exhausted and the star will evolve directly to
become a hybrid white dwarf having a carbon-oxygen core with a helium
cocoon.

Providing the hydrogen envelope is sufficiently massive, core
contraction after helium exhaustion causes the hydrogen shell to heat
and returns the star to the Asymptotic Giant Branch (AGB), where it
develops a degenerate carbon-oxygen core. Thermal pulses are
associated with unstable burning in the helium shell, whilst the
star's great luminosity comes from the hydrogen shell.  Convective
dredge-up, strong stellar winds and interaction with potential binary
companions produce a variety of stellar types associated with the AGB.
Further details are given by \cite{2005ARA&A..43..435H}.

Ultimately, the hydrogen layers of an AGB star are blown away, and
hydrogen-burning ceases.  With no nuclear support, the star contracts
rapidly, first at constant luminosity, initially as an 
{\bf embedded  IR source}, then as a {\bf post-AGB F or A star}, 
then as a {\bf subdwarf O star}.  The remnant of the hydrogen envelope
may still be close enough to be ionised by the hot
{\bf central star of a planetary nebula (CSPN)}.

At this point the star is almost a {\bf carbon/oxygen WD}, with a thin
helium shell and an even thinner hydrogen envelope.  The mass of the
helium shell is dictated by the point of the AGB thermal-pulse at
which hydrogen-burning switched off.  With the right conditions, the
unstable helium shell my make one final shell flash whilst the stars
is contracting, or even after it has reached the white dwarf cooling
track. This final flash forces the star to expand in just a few years,
mixing the outer layers with massive amounts of helium, carbon and
oxygen. After expansion, the star again contracts, possibly as a 
{\bf hydrogen-deficient CSPN}, and later a {\bf PG1159 star} or 
{\bf DO white dwarf}.
 
If the expansion of a normal giant or AGB star brings it into contact
with a less massive companion, a sudden transfer of mass can produce a
common-envelope surrounding both stars. Friction makes the stars lose
orbital energy and spiral towards one another, whilst the heat
generated expands and expels the common envelope. The stripped giant
may now be, for example, a white dwarf in a close binary system.  When
the secondary itself becomes a giant, the process can be repeated,
yielding a hot subdwarf or another white dwarf.  By this and other
means, binary systems can evolve to the point where both components
are WDs.

Theory suggests that all binaries will emit gravitational radiation
(GR) at a rate proportional to $P^{-2/3}$. Whilst GR removes orbital
energy, it is only significant for very short-period binary white
dwarfs\footnote{The GR timescale for a CO+CO WD binary with a period of
 2 hr is approximately $10^{10}$ yr.}. When the orbit has shrunk so
that the less massive component fills its Roche lobe (and the period
has been reduced to 3 minutes or so), mass transfer will
commence. Loss of mass causes the WD donor to expand. Transfer of mass
causes the orbit to expand.  Several outcomes are conjectured. If the
mass ratio (donor/receiver) is small, the donor expansion will be
slower than the orbit expansion, so mass transfer is stable, and
possibly intermittent. Likely candidates are {\bf AM\,Canem
Venaticorum (AM\,CVn) variables}, short-period
hydrogen-deficient white dwarf binaries with an accretion disk.  If the
mass ratio is closer to unity, the expansion of the donor exceeds the
orbital expansion, so a runaway process occurs, the donor is destroyed
on the timescale of an orbital period and the debris accretes on to
the receiver.  What happens next depends on the WD masses.  The merger
of two helium WDs is thought to lead to helium ignition and the
creation of a hot subdwarf, possibly a {\bf helium-rich sdB or sdO
 star}. This may explain the fraction of sdB stars believed to be
single. A CO+He WD merger will lead to helium-shell ignition, and has
been argued to produce {\bf R\,Coronae Borealis (RCB) variables} and
{\bf extreme helium (EHE) stars}. If the final combined mass
approaches the Chandrasekhar limit, a type Ia supernova may ensue.

%% csj -- I haven't touched the LMXB stuff 

A massive star will not develop a degenerate core and helium burning
begins without the Helium Flash. Eventually a supernova explosion is
expected to occur and if in a binary, and the binary remains bound, a
{\bf low-mass X-ray binary (LMXB)} could result where one component is a
neutron star (or black hole) and the other an unevolved object often
referred to as the donor star. Modelling and understanding the
survival of a binary during a supernova explosion presents a
challenge, as does the subsequent transfer of mass from the donor star
to a neutron star or black hole.

\subsection{Gaia and fundamental parameters}

While the formation and evolution of a few of the objects
representative of the late stages of stellar evolution appear to be
understood in principle, there is an uncomfortable dependency on
luminosities inferred from companions in binary systems where it is
supposed that the companion has evolved as though it were a single
star.  None of the types of object briefly discussed above is close
enough to the Sun to have allowed an accurate parallax measurement
from the ground or by the Hipparcos satellite. Gaia can be expected
to provide accurate distance determinations for most of the various
types of object discussed above.

Accurate luminosities follow from precise distances, and from these,
stellar energy distributions at the tops of the photospheres may be
directly inferred from their observed counterparts at the top of the
Earth's atmosphere; this provides an exacting test of model stellar
atmosphere calculations, which is particularly important for hot stars
where Local Thermodynamic Equilibrium (LTE) cannot be assumed and
line-blanketting is necessarily more approximate. The likely need for
improved non-LTE model stellar atmospheres would stimulate new atomic
physics calculations of radiative and collisional data. Any
improvement to model stellar atmospheres should give better
determinations of effective temperature, surface gravity and
abundances. Comparisons of improved abundances are needed for
studying the consequences of mass transfer in binary systems. Better
stellar radii follow from accurate luminosities and improved effective
temperatures.

With a better knowledge of luminosities and radii, stellar evolution
models are more constrained. In the context of hot stars in a poorly
understood late stage of stellar evolution and known to be a component
of a binary system, a knowledge of the luminosity in particular will
facilitate an understanding of differences between binary and single
star evolution. The evolution of some objects, which are apparently
single stars, can only be understood if they formed from the merger of
two white dwarfs; precision measurements of parallax, proper motion
and radial velocity by Gaia will help confirm these as single objects
and so probably a consequence of a merger between two WDs. In the
case of sdB stars, for example, it would then be possible to make a
detailed comparison between those formed by a WD merger and those
formed as a consequence of binary interactions affecting RGB
evolution.

The adequacy of stellar evolution calculations for interpretting
colour-magnitude diagrams of resolved stellar populations is reviewed
by \citet{2005ARA&A..43..387G}; they emphasise the need for a better
knowledge of conductive opacities, neutrino energy losses and the 
correct nuclear cross-section for the 
$^{12}{\rm C}(\alpha,\gamma)^{16}{\rm O}$ reaction.
The calculated size of the helium core at the top of the Red Giant 
Branch, which determines the luminosity at this point, the subsequent
luminosity on the Horizontal Branch, core helium burning lifetime and
understanding the subsequent stages of stellar evolution directly 
depend on our knowledge of conductive opacities, neutrino energy losses
and the correct nuclear cross-section for the 
$^{12}{\rm C}(\alpha,\gamma)^{16}{\rm O}$ reaction.  At the moment
stellar evolution models fit observed colour-magnitude diagrams of
resloved stellar populations reasonably well but Gaia can be expected
to provide observations as outlined above which place much tighter
constraints on the models.

Accurate parallaxes, proper motions and radial velocities measured by
Gaia would give reliable space motions relative to the Local Standard
of Rest of all stars observed and for those representing the poorly
understood late stages of stellar evolution in particular.  A
knowledge of space motions then enables the stellar evolution of the
objects concerned to be studied in the wider context of the dynamical
evolution of the Galaxy.  It would for example be possible to
establish whether helium-rich subdwarf-O stars have a different
galactic distribution from their helium-poor counterparts.

Although much can be anticipated using Gaia photometry,
spectrophotometry from the far ultraviolet to far infrared would 
also be needed for an exacting test of model stellar atmospheres.
Much useful spectrophotometry, or measurements of flux density as a
function of wavelength already exists in archives from previous space
missions and ground-based observations; prospects for obtaining more
such data already exist and could be proposed for future space missions
and ground-based observatories.  For binary stars, the prospect of 
time-dependent observations of flux density is eagerly anticipated
through multi-wavelength stellar astrometry, to be obtained by the proposed
Sim Lite mission \citep{2010ApJ...717..776C}.

\subsection{Gaia and hot subdwarfs}

Hot subluminous stars \citep{2009ARA&A..47..211H}
are an important population of faint blue stars
at high Galactic latitudes and are immediate progenitors to white
dwarfs.  They have recently been studied extensively because they are
common enough to account for the UV excess observed in early-type
galaxies. Several thousands have been identified in the Galaxy and
outnumber the white dwarfs among the faint blue stars ($U-B < -0.4$)
to 18 mag. Pulsating sdB stars became an important tool for
asteroseismology, and sdB stars in close binaries may qualify as
Supernova Ia progenitors.

Subluminous B stars (sdB) have been identified as extreme horizontal
branch (EHB) stars, {\it i.e.} they are core helium-burning stars with
hydrogen envelopes that are too thin to sustain hydrogen burning.
Therefore they evolve directly to the white-dwarf cooling sequence by
avoiding the AGB. The fraction of sdB stars
in short period binaries is very high, about half of the sdBs
are found in binaries with periods of ten days or less.  Obviously,
binary evolution plays an important role in the formation of sdB
stars, in particular when merging of close binary white dwarfs system
are also considered to explain the formation of single stars.

However key issues in the field remain to be resolved:

\begin{itemize}

\item[1)] The most fundamental property of a stars is its
  mass. Because no direct measurement is available for a hot subdwarf
  star, it is often assumed that an sdB star has half a solar mass,
  which is the canonical mass for the core helium flash. Binary
  population synthesis predicts mass distributions from 0.3 to 0.8
  M$_\odot$. The details depend on several poorly constrained
  parameters. Asteroseismology has successfully probed the
  interior of sdB stars and determined masses and envelope masses, but
  the results are model dependent.

\item[2)] Most of the companions to binary sdB stars are very faint and
  are therefore outshone by the primary. From indirect mass estimates a
  large variety of companions have been identified, ranging from
  substellar objects (brown dwarfs) and low-mass main sequence stars
  to white dwarfs and even more massive compact objects (possibly
  neutron stars). Usually, only the mass ratio of the system can be
  determined and, therefore, the interpretation is hampered by
  the unknown mass of the sdB star.
  
\item[3)] The stellar population of the Galactic bulge is similar to
  that of early-type galaxies and, therefore, provides a laboratory
  to study the UV-excess phenomenon. Hot subluminous B have been
  discovered in the Galactic bulge but no quantitative results are
  available yet.

\item[4)] Space densities, scale heights and birthrates of the disk
  sdB stars are poorly constrained. Subluminous B stars have an
  absolute visual brightness similar to the Sun with little scatter:
  M$_{\rm V} = 4.2 \pm 0.5$ mag. Hence they sample both the thin and thick
  disk. An accurate knowledge of the scale heights are required to
  derive the birthrates, which is a key to understanding the various
  evolutionary channels that lead to the formation of a white dwarf.

\item[5)] Hot subluminous stars can be traced out into the
  halo. Indeed, halo sdBs have been identified in galactic globular
  clusters as well as in the field.  Comparing the different
  population of sdB stars (thin disk, thick disk and halo) will allow
  the time and metallicity dependence of sdB evolution to be studied.

\item[6)] The fastest moving stars allow the dark matter
  mass of the galactic halo to be probed. Radial velocity surveys 
  revealed the
  existence of hyper-velocity stars (HVS) travelling so fast that
  they may be unbound to the Galaxy. Amongst the first three stars
  discovered serendipitously in 2005, the hot subdwarf US~708 stands
  out as the only low mass, highly-evolved HVS known. The place of
  origin of HVS is supposed to be the Galactic centre, because it
  hosts a supermassive black hole (SMBH) and tidal disruption of a
  binary by a SMBH is believed to be the only mechanism capable of
  achieving the required acceleration.

\item[7)] Subluminous B stars in close binaries are very
  frequent. However, surveys have been restricted to periods of ten
  days or less. The frequency of longer period systems is 
  poorly constrained.

\end{itemize}

Gaia will resolve these issues. Items 1 and 2 will be resolved through
accurate parallaxes. Items 3 to 6 will require the accurate proper
motions in addition to parallaxes. At $V$=20 mag, Gaia will measure and
discover lots of hot subdwarfs in Baade's window, in the galactic
bulge and in the halo (to 20 kpc), and also in the Galactic Plane,
which is still largely {\it terra incognita} as UV excess surveys have had to
avoid the Galactic Plane.  The RVS is an ideal instrument to tackle
item 7. Although the spectral window around the CaT lines is not well
suited for early type stars, sdB stars do show a couple of lines
suitable for radial velocity measurement. The five years of operation
will allow a search for long-period radial velocity variations. A
sufficiently large sample of sufficiently bright sdB stars ($<$17
mag) is available to derive statistically significant results.

%\subsection{Gaia, planetary nebulae and their central stars}
%Ralf$
\subsection{Gaia and cataclysmic variables}

Cataclysmic variables (CVs) are close interacting binaries in which a
white dwarf accretes from a low-mass companion star, and represent an
important benchmark population for the study of a wide range of
astrophysical questions.

These include, on the one hand, the evolution of binary stars. CVs
descend from main sequence binaries with unequal mass ratios. As the
more massive star evolves off the main sequence, it eventually fills
its Roche volume, initiating mass transfer onto its lower-mass
companion. This mass transfer is unstable in most cases, and leads to
a common envelope phase. Following the common envelope, the evolution
of CVs is driven by orbital angular momentum loss via stellar magnetic
wind braking and/or gravitational wave radiation.  Common envelope
evolution and angular momentum loss are critical ingredients in the
formation of a vast range of exotic objects, such as stellar black
hole binaries, milli-second pulsars, or double-degenerate white dwarf
pairs and neutron star pairs, which are potential progenitors of
SN\,Ia and short GRBs, respectively. Despite the overarching
importance, our understanding of the physical processes that are at
work in the evolution of compact binaries are still in its infancy,
and the observational constraints are poor. Because they are common,
easily accessible to ground and space based studies, and structurally
simple in terms of their stellar components, CVs are the natural
choice for observational population studies that can test and guide
the further development of the theories of compact binary evolution.

Beyond these bulk properties, CVs are excellent laboratories for
accretion physics. Most CVs contain a non-magnetic white dwarf, and
the mass flow occurs via an accretion disc. The discs in CVs share
many physical properties with the discs in X-ray binaries,
proto-stellar systems, and AGN, but offer the advantage that their
structure can be directly probed by orbital phase-resolved
observations, either via Doppler tomography 
\citep{1988MNRAS.235..269M} or eclipse mapping 
\citep{1985MNRAS.213..129H}.
The small physical size of CV discs
also implies that their dynamical behaviour can be studied in exquisite
detail on time scales of weeks to months 
\citep{2001NewAR..45..449L,2010ApJ...725.1393C}.

\begin{figure}
%\centerline{\includegraphics[angle=-90,width=14cm]{cv_uggr_ggr.ps}}
\includegraphics[angle=-90,width=14cm]{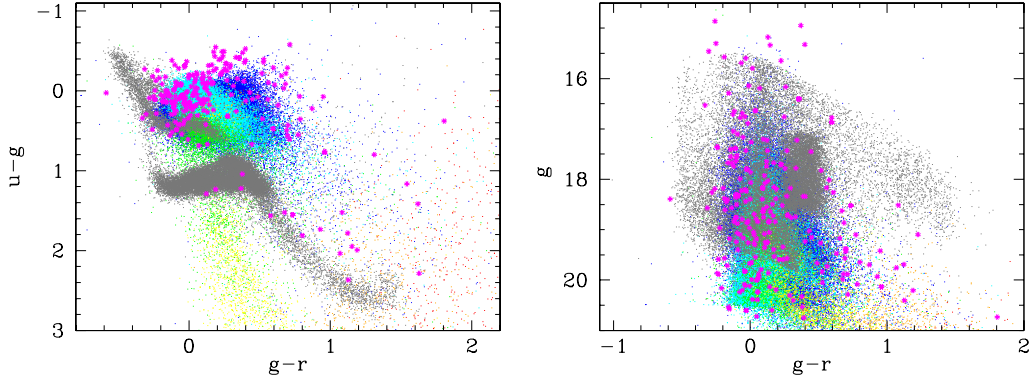}
\caption[Population of the CVs discovered by SDSS]{\label{f-cvcolours} The population of the CVs (purple stars)
  discovered by SDSS in a $(u-g,g-r)$ colour-colour diagram (left) and
  $(g,g-r)$ colour-magnitude diagram (right). Within SDSS, CVs overlap
  very strongly with the parameter space spanned by quasars (shown
  with increasing redshift as blue, cyan, green, yellow, orange and
  red dots). Main-sequence stars and white dwarfs are shown in
  gray. Gaia will provide comparable all-sky colour information, as
  well as distances, proper motions, and variability, unambiguously
  identifying all CVs to its magnitude limit.}
\end{figure}

\textit{The impact of Gaia}

\begin{itemize}
\item In the past, observational population studies of CVs were
  haunted by selection effects, leading to severely skewed and
  incomplete CV samples \citep[e.g.][]{1999MNRAS.309.1034K}. Over the
  past few years, this situation has changed markedly as SDSS has
  identified a large and homogenous sample of CVs that led to new
  constraints on the theory of CV evolution 
  \citep{2009MNRAS.397.2170G,2011arXiv1102.2440K}.
  SDSS efficiency in finding CVs was a by-product
  of their non-stellar colours, which mixes them into the parameter
  space spanned by quasars (Fig.\ref{f-cvcolours}). Gaia will take
  observational CV population studies to an unprecedented level of
  detail, as the combination of colour, distance, proper motion, and
  variability is bound to identify every single CV down to the Gaias
  magnitude limit. 

\item CVs are intrinsically faint, and, compared to single white
  dwarfs, relatively rare. Hence, Hipparcos observed only a tiny
  handful of CVs, and despite additional intensive (and expensive!)
  ground and space-based parallax programs, the number of CVs with
  accurate distances remains very small 
  \citep{2003AJ....126.3017T,2008AJ....136.2107T,
   2000AJ....120.2649H,2003A&A...412..821B}. Gaia
  will measure high-quality parallaxes of pretty much all known (and
  new) CVs. The knowledge of the distances to many hundreds of CVs
  will truly revolutionise the field, as it will lead to accurate
  determinations of the mass transfer rates, a key parameter both for
  the evolution of the systems, as well as for the structure and
  dynamics of their accretion discs. In addition, it will be possible
  to measure the white dwarf masses of a large number of CVs (based on
  the Gaia distances, spectrophotometric radii, and mass-radius
  relations), which will settle the important question whether the
  accreting white dwarfs grow in mass in CVs grow, or not.
\end{itemize}

\textit{Data needs beyond Gaia.}

\begin{itemize}
\item A detailed classification of the CVs discovered by Gaia will
  require spectroscopic follow-up at higher spectral resolution
  ($R\sim5000$) covering the entire optical range ($3800-9200$\,\AA). 
\item The Gaia scanning law is too coarse to provide accurate
  long-term light curves of CVs, denser ground-based monitoring, such
  as currently carried out by e.g. the Catalina Real Time Transient
  Survey \citep{2009ApJ...696..870D} would be highly desirable. 
\item Gaia will discover a large number of eclipsing CVs, which are
  key targets for detailed physical parameter studies. This will
  require access to 2--4\,m telescopes in both hemispheres, equipped
  with high-speed CCD cameras such as ULTRACAM 
  \citep{2007MNRAS.378..825D}.
\end{itemize}

\subsection{Gaia and AM CVn variables} 

AM~CVn stars are interacting binary stars of extremely short period 
(5 to 65 minutes) in which white dwarfs accrete from 
hydrogen-deficient, degenerate companions which may themselves once
have been white dwarfs. They, and the related but much more numerous 
detached double white dwarfs, are of interest as gravitational wave 
sources and possible Type~Ia supernova progenitors. The properties of
AM~CVn stars depend strongly on their orbital period. At short periods
($P < 20$ mins) they have bright accretion discs which outshine the
underlying white dwarf. The systems are in a permanent high state;
examples are AM~CVn itself and the recent discovery from Kepler,
SDSS~J1908+3940 \citep{2011ApJ...726...92F}
At intermediate periods, $20 < P < 40\,$mins, AM~CVn stars show 
outbursts of a few magnitudes, while for $P > 40\,$mins they seem to
exist in a stable low state, although they can exhibit considerable
short-timescale flickering activity.

AM~CVn stars are hard to find, and we know of only $\sim 30$ systems,
and have measured periods of only 18 of these. The best current
estimate of their space density is from 1 to 
$3\times 10^{-6}\,\mathrm{pc}^{-3}$ \citep{2007ApJ...666.1174R}.
If all had absolute magnitudes similar to the 
faintest so far observed, $M_V = 12$, we could expect from 200 to 600 
AM~CVn stars to be surveyed by Gaia down to $V = 20$. AM~CVn stars can
however be considerably brighter than $M_V = 12$, (several are known
with $M_V = 5$ to $7$, \citeauthor{2007ApJ...666.1174R})
so several thousand may be within reach.

Gaia can help in the discovery of AM~CVn stars in the following ways: 
(i) location in the HR diagram as a population of blue, faint objects,
(ii) variability, (iii) colours and (iv) RVS spectra. The first two of
these will end with AM~CVn stars being grouped with other faint blue
variables, with pulsating sdBs and DB white dwarfs overlapping at the
bright and faint ends of their distribution, but with hydrogen-rich
cataclysmic variable stars (CVs) likely to be the major source of
confusion since they have very similar variability characteristics,
colours and magnitudes.  The chief means to select between AM~CVn stars
and CVs from Gaia data alone will be methods (iii) and (iv). Colours
will differ in detail because of the strong He~I lines of long period
AM~CVn stars (the majority) compared to the strong Balmer lines of CVs.
The RVS spectra may also provide some discrimination from the presence
of Ca~II and Paschen lines in CVs versus N~I lines in AM~CVns in the
range the RVS samples. The separation is unlikely to be perfect, 
and ground-based follow-up will be essential to confirm AM~CVn stars
from Gaia, as it will for determination of their orbital periods.
Nevertheless, there are excellent prospects for a significant addition
to the current small sample, and of course Gaia will provide us with
secure distances for these poorly understood systems.

\subsection{Gaia and low-mass X-ray binaries}
Black holes and neutron stars provide the best laboratories in the
Universe for studying general relativity.  Stellar mass black holes
represent an ideal class of system for understanding accretion onto
black holes -- the vast majority of accretion physics is "scale-free",
meaning that it should function in the same manner for stellar mass
black holes in X-ray binaries as it does for the supermassive black
holes in active galactic nuclei -- and without the solid surface that
provides complications in the cases systems with neutron star and
white dwarf accretors.  On the other hand, the viscous timescale from
the inner edge of the accretion disk in most active galactic nuclei
will be longer than the lifetime of the typical astronomer -- AGN
variability is fundamentally "weather", while in X-ray binaries one
can understand how systems respond to changes in mass transfer rate.

Accreting neutron stars present their own opportunities to understand
fundamental physics.  The equation of state of neutron stars
determines their mass-radius relation, and is determined by
interactions between particles at nuclear density.  Accurate
measurements of the masses and radii of neutron stars thus give the
opportunity to study particle interactions in a regime which cannot be
probed in a laboratory or an accelerator.  It has also been suggested
that neutron stars' spin evolutions may be controlled, in part, by
gravitational radiation.

Finally, black holes and neutron stars are the remanants of the most
massive stars -- the stars which provide most of the metal enrichment
in the Universe.  Understanding their key system parameters -- their
masses, distances, space velocities, and orbital inclination angles --
is fundamentally important for understanding how these objects form,
and for taking full advantage of their capability of probing
relativity and accretion physics.

Gaia will revolutionise a few key aspects of studies of neutron stars
and black holes.  Both distances and proper motions of X-ray binaries
are necessary for extracting some of the most important fundamental
physics that can, in principle, be obtained from these systems.
Additionally, searches for astrometric wobble associated with black
holes in wide binaries would be fundamentally important for
understanding the black hole mass function.

\subsubsection{Distances}

Accurate distance measurements to X-ray binaries are notoriously
difficult.  At the present time, two geometric parallax measurements
have been reported, both based on radio data (Sco X-1 --\citealt*{bradshaw99}; and V404 Cyg). %-- \citealt{miller99}). 
 About 10\% of X-ray binaries are located in globular
clusters, so their distances are known to about 10\% accuracy as well.

The bulk of X-ray binaries have had their distances estimated by far
more indirect techniques -- for typically from using the flux and
temperature of the donor star, plus an estimate of the radius of the
donor star based on the requirement that it fills its Roche lobe at
the measured orbital period.  Unfortunately, in most X-ray binaries,
even in quiescence, the accretion light contributes significantly in
the optical and infrared, leading to difficulties in measuring the
actual flux of the donor star properly.  This leads to systematic
uncertainties in the distances to the systems.

It also leads to incorrect interpretations of the size of the
ellispoidal modulations of the donor stars, which are used to estimate
the inclination angles, since the donor stars' ellipsoidal modulations
are being diluted by the accretion light.  Direct distance
measurements from parallax thus can also be used to decompose the
light into donor star and accretion light components.  This in turn
will give better mass estimates for the accretors, and allow spectral
fitting of the remaining accretion light in quiesence allowing for
better modeling of accreting black holes and neutron stars at the
lowest luminosities.

For the neutron star systems in particular, getting accurate distances
is of great importance.  The radii of neutron stars can be estimated
by fitting models to their X-ray spectra in the two cases where they
are dominated by thermal emission from the surface of the neutron star
-- in quiescence, or during the Type I X-ray bursts -- the runaway
nuclear burning episodes that take place on the neutron star surface.
However, the radius measurements are only as good as the distance
estimates to the neutron star.  While this has been done in some cases
in globular clusters, it is of great value to have as large a range of
neutron stars as possible to test, since a single combination of mass
and radius for a single neutron star can, in principle, rule out a
wide array of equations of state for supranuclear density matter.

{\it Observable systems}

X-ray binaries are among the rarest binary stars in the Galaxy.
Still, approximately 25 X-ray binaries are brighter than 20th
magnitude in $V$, including several black hole systems (GRO J1655-40,
V404 Cyg, GX 339-4, Cygnus X-1, XTE J1118+480) -- see Liu et al.  for
a compilation of X-ray binaries with properties including their donor
star fluxes.  It is additionally likely that some quiescent X-ray
transients will be discovered through Gaia observations.  The orbital
period distribution of X-ray binaries is extremely difficult to
estimate either from theoretical concerns (because of the large number
of poorly constrained parameters that go into binary evolution
calculations) or from observations (because to date, nearly all known
X-ray binaries were first discovered in outburst, leading to
potentially strong selection effects, since the probability that a
source will have undergone an outburst over the 50 year history of
X-ray astronomy is likely to be strongly correlated with its orbital
period).  Nearly any X-ray binary will have a radial velocity
amplitude larger than 10 km/sec, and so it seems likely that Gaia will
detect nearly all X-ray binaries brighter than $V=17$.

{\it Inclination angles}
Getting good inclination angle estimates is important for a few other
reasons.  Several systems seem to show evidence of relativistic jets
which are not perpendicular to their binary orbital planes, which
could be interpreted as evidence that the black hole spins in these
binaries are misaligned \citep{maccarone02}.  Better distance estimates
will allow for the relativistic motions of the jets to be converted
into inclination angles and jet speeds more accurately, while better
estimates of the inclination angles of the binary planes will allow
firmer conclusions about the misalignment of jets, perhaps leading to
strong evidence for rapidly rotating black holes, derived in a manner
completely independent from the normal X-ray spectral fitting methods.
Additionally, the GEMS mission, soon to be launched by NASA, will
attempt to find evidence for rotating black holes from the X-ray
polarisation properties of thin accretion disks around black holes --
but the interpretation of these measurements, too, will be inclination
angle dependent.

\subsubsection{Proper motions}

Isolated radio pulsars show a space velocity distribution far larger
than that of their parent population or massive stars, indicating that
they are often, or perhaps always, formed with strong natal kicks.
The evidence for such kicks at kick for neutron stars in X-ray
binaries, and especially for black holes, is less secure.  About a
decade ago, it was generally believed that neutron stars would all
form with kicks, while black holes would not.  The spin
period-eccentricity relation in double neutron stars \citep*{dewi02}
 and the high space velocity of the black hole X-ray binary
GRO J1655-40 have cast doubt on this sharp distinction.  Theoretical
work done in recent years has also suggested that electron capture
supernovae might be responsible for the formation of a low velocity
kick mode of neutron star formation, while some black holes might form
through a process in which a supernovae takes place forming a neutron
star which lives only a short time before a fall-back disk accretes
onto it, producing a black hole.  It is thus of great interest to
measure the proper motions of X-ray binaries, in conjunction with
their distances, to determine whether they have locations and
velocities consistent with having been formed with velocity kicks.

\subsubsection{Direct inclination angle measurements}

For a few systems, Gaia may be able to map out the orbit of the donor
star, leading to a direct measurement of the system inclination angle
(under the well-justified assumption of a circular orbit).  Such
measurements will provide an extra degree of redundancy which will
allow testing of the ellipsoidal modulation method of inclination
angle estimation.  They will additionally give the position angles of
the orbits, which will allow for additional checks on whether jets are
perpendicular to the binary orbital planes.

\subsubsection{Non-accreting binaries with black holes and neutron stars}

At the present time, with the exception of a few microlensing black
hole candidates, the black holes known in the Milky Way are all in
accreting binary systems.  The masses of the black holes in these
binaries are almost certainly affected by binary evolutionary
processes.  On the other hand, most massive stars are not in such wide
binaries.  The mass function of black holes measured to date is thus
severely biased in ways which are not well understood.

Gaia holds the potential to help solve this problem, by indentifying
very wide binaries containing compact objects through either
astrometric wobble, or radial velocity wobble.  About $10^8-10^9$
black holes should have formed over the lifetime of the Galaxy \citep{heuvel92}.
  Assuming that the space density of black holes
traces the space density of stars in general, there should then be
approximately $10^3-10^4$ black holes within 100 pc.  Within that
distance range, even M dwarf counterparts should be detectable by
Gaia, although relatively few white dwarfs will be detectable.  

The optically emitting star in a wide binary containing a black hole
will be a small fraction of the total mass of the binary system
(except in the cases of very bright, massive stars).  As a result the
motion of the optical star will account for nearly the full orbital
motion of the binary, so 1 milliarcsecond wobble will result for
orbital periods as short as about 2 weeks, even at a 100 pc distance.
Systems with orbital periods shorter than 2 weeks will have radial
velocity amplitudes of order 100 km/sec or faster, so all binary
systems containing black holes will be detected by Gaia as either
spectroscopic or astrometric binaries unless they are almost perfectly
face-on short period binaries, or have orbital periods sufficiently
long that the astrometric orbital motion shows no curvature on the 5
year mission it.  The orbital period distribution of binaries
containing black holes is almost completely unknown, but if one
assumes that it follows the distributions followed by G dwarfs
\citep{duquennoy91}, that of a log normal distribution peaking
at 180 years, with a disperison of 2.3 dex, we find that about 1/3 of
the binaries should have periods shorter than about 20 years.  It thus
seems reasonable to expect that at least 100 or so wide binaries with
black hole primaries will be discovered by Gaia, with around 1/3 of
them having orbital periods shorter than 2 years, thus allowing for
Gaia to have sampled the orbit directly, thus giving the opportunity
to measure precise black hole masses -- this sample would be similar
in size to the sample of stellar mass black holes in X-ray binaries
whose masses have been estimated, allowing for a direct probe of
whether binary evolutionary strongly affects the black hole mass
function.

\subsection{Gaia and White Dwarf Mergers}

The dynamical mergers of two white dwarfs are believed to give rise to
progeny of several types depending on initial masses and
composition. The most populous are likely to be low-mass helium
main-sequence stars, most likely manifest as {\bf helium-rich sdB
  or sdO} stars, or possibly as {\it single} {\bf subdwarf B} stars. 
The next most populous are the {\bf R\,CrB variables} and {\bf
  extreme helium stars}. More massive mergers are conjectured to
result in an explosion of one type or another, either {\bf 
Type Ia SNe} or {\bf short-duration $\gamma$-ray burts}. 
A number of critical questions will be answered directly using Gaia
6-space measurements:
\begin{itemize} 

\item[]{\it What is the galactic distribution of double-white dwarf
  binaries (DWD's)?} Binary-star population-synthesis studies provide quite
  specific predictions regarding the space density and distribution 
  of DWD's as a function of type and history. In particular, these
  predictions are sensitive to the star formation history of the
  Galaxy (including rate, and initial mass and period distribution), 
  and to assumptions about the physics of common-envelope 
  ejection in close binaries. Contemporary surveys are 
  discovering increasing numbers of short-period DWDs. Complete 
  six-space positions for all observable DWDs will verify these assumptions. 

\item[] {\it Is there a demonstrable connection between the
  galactic distribution of double white dwarfs and their supposed
  progeny?} For example, from their distribution on the sky, 
  R\,CrB and EHe stars appear to belong to a
  galactic bulge population. However population synthesis studies
  suggest that their binary progenitors must come from a much younger 
  galactic disc population. With accurate distances and proper
  motions for RCBs and EHEs (for example), 
  this question will be resolved immediately. 

\item[]{\it What are the masses and luminosities of supposed merger
  products?} Binary-star population-synthesis models predict the
  mass-distribution of DWD mergers. Models for post-merger evolution 
  predict evolution tracks as functions of merged mass. At present,
  luminosites and masses can only be inferred. Direct distance
  measurements will give both and will verify conclusively
  the post-merger evolution calculations. They will also test whether 
  DWD mergers are likely to be conservative (in mass), and hence the
  physics of the merger process. 

\item[]{\it Do DWD mergers produce explosions?} Some models predict
  that Type Ia supernovae and possibly some short-duration
  $\gamma$-ray bursts arise from super-Chandrasekhar and some 
  He+He mergers respectively. Comparing space densities of DWDs with 
  observed Ia and GRB rates will address this question.

\end{itemize}

\putbib[late_refs]

%% file: section_projects.tex
\newpage
\section{Synergy with future projects}

\subsection{Provide candidates for Plato}

In the framework of the PLATO project, the 1st option to built the PLATO Input 
Catalogue (PIC), from which to select the PLATO targets, is to use a 
preliminary Gaia catalogue. 
Specific informations about this issue can be found in several PSPM (PLATO 
Science Preparation Mangement) documents, e.g. from
http://www.oact.inaf.it/plato/PPLC/PSPM/PSPM.html.
\\
This general choice has important implications also for the white dwarfs: 
the large sample of white dwarfs that will be observed by Gaia
($\approx$400,000 objects, \citealt{jordan2007})
%
% ... MAY BE THERE IS A MORE RECENT ESTIMATE ??
%
will contain a $\sim$3-4\% subsample of DAV pulsators (plus a much smaller 
fraction of DBVs and PG~1159 pulsators).
For some of them, pulsations will be directly detected by Gaia photometry
(see Sect.\,\ref{sec:directperiod} for more details).
For the others, Gaia will at least tell us precise distances and luminosities,
from which it will be possible to isolate those close to the DAV instability 
strip.
\\
The possibility to observe some of these pulsators with PLATO is an interesting
development considering that it is unlikely that these objects will be observed
by CoRoT or Kepler.
The probability to find a sufficiently bright WD pulsator in its field of 
view is virtually zero for CoRoT and very small for Kepler.
Moreover, as the Kepler ``survey phase" (in which the best asteroseismic 
targets have been selected) is already finished, it is quite unlikely that a 
new WD pulsator in the Kepler FOV can be identified and observed by Kepler
in the next months/years of the mission.
\\
On the other hand, the number of DAV pulsators in the much larger PLATO FOV 
will be large enough to be able to observe some of them.
Considering a limiting magnitude $V\le$16.0 and a local space density of DAVs 
of 1.9$\times$10$^{-4}$ pc$^{-3}$ (obtained from the number of DAVs in the 
local WD sample of \citealt{holberg+2008}, see also \citealt{sion+2009}), 
the number of DAVs in a single PLATO FOV of $\sim$2,200 sq. degr. should be 
about 10.
Considering that PLATO will observe different fields (2 long-term fields of 
2-3 yrs each plus another 5-6 step \& stare fields of 2-5 months each), the 
total number of DAVs observable with PLATO is about 70.
These numbers increase by a factor of 2 when going deeper by half magnitude.
\\
In conclusion, a timely identification of white dwarfs in the Gaia catalogue, 
possibly followed by ground-based spectroscopy/photometry in order to select 
the best DAVs or DAV candidates, is an important issue in order to be able to
observe some of them with PLATO.
Thanks to the outstanding quality of PLATO data in terms of duty cycle, 
photometric accuracy and homogeneity, these observations would be a major step
towards detailed WD asteroseismology, allowing to study in much greater detail 
the internal structure of these degenerate stars.

\putbib[projects_refs]